\documentclass[twocolumn,superscriptaddress,amsmath, amssymb, amsfonts,preprintnumbers,aps,prd,longbibliography,nofootinbib]{revtex4-2}

\usepackage{verbatim}
\usepackage{hhline}
\usepackage{graphicx}
\usepackage{bm}
\usepackage{amstext}
\usepackage[table,dvipsnames]{xcolor}
\usepackage[caption=false]{subfig} 
\usepackage{amsmath}
\usepackage{bbold}
\usepackage{delimset} 
\usepackage[colorlinks=true]{hyperref}
\usepackage{physics}
\usepackage{empheq}
\usepackage[normalem]{ulem}

\usepackage{orcidlink}

\makeatletter

    \def\CT@@do@color{%
      \global\let\CT@do@color\relax
            \@tempdima\wd\z@
            \advance\@tempdima\@tempdimb
            \advance\@tempdima\@tempdimc
    \advance\@tempdimb\tabcolsep
    \advance\@tempdimc\tabcolsep
    \advance\@tempdima2\tabcolsep
            \kern-\@tempdimb
            \leaders\vrule
                    \hskip\@tempdima\@plus  1fill
            \kern-\@tempdimc
            \hskip-\wd\z@ \@plus -1fill }
    \makeatother

\DeclareFontFamily{OT1}{pzc}{}
\DeclareFontShape{OT1}{pzc}{m}{it}{<-> s * [1.350] pzcmi7t}{}
\DeclareMathAlphabet{\mathpzc}{OT1}{pzc}{m}{it}

\def\cO{\mathcal{O}}

\def\eps{\epsilon}

\def\d{\mathrm{d}}

\def\d{\mathrm{d}}
\def\eps{\epsilon}

\def\nn{\nonumber}

\def\Eqn#1{Eq.~\eqref{#1}}

\def\Fig#1{Fig.~{\ref{#1}}}

\def\Tab#1{Table~{\ref{#1}}}

\def\App#1{Appendix~{\ref{#1}}}

\def\Rcite#1{Ref.~\cite{#1}}
\def\Rcites#1{Refs.~\cite{#1}}

\newcommand*{\vct}[1]{\boldsymbol{#1}}
\newcommand{\sig}{\sigma}

\newcommand{\be}{\begin{equation}}
\newcommand{\ee}{\end{equation}}
\newcommand{\ba}{\begin{align}}
\newcommand{\ea}{\end{align}}
\ifx\genfrac\sdflkaj\else\fi
\newcommand{\sfrac}[2]{{\textstyle\frac{#1}{#2}}}

\newcommand{\pin}{p_\infty}

\newcommand{\gam}{\gamma}
\newcommand{\Gam}{\Gamma}
\newcommand{\Del}{\Delta}
\newcommand{\mO}{\mathcal{O}}

\begin{document}

\preprint{HU-EP-24/07-RTG}

\title{Post-Minkowskian Theory Meets the Spinning Effective-One-Body Approach\\ for Two-Body Scattering}

\author{Alessandra Buonanno\,\orcidlink{0000-0002-5433-1409}}
\email{alessandra.buonanno@aei.mpg.de}
\affiliation{Max Planck Institute for Gravitational Physics (Albert Einstein Institute), Am M\"uhlenberg 1, 14476 Potsdam, Germany}
\affiliation{Department of Physics, University of Maryland, College Park, MD 20742, USA}

\author{Gustav Uhre Jakobsen\,\orcidlink{0000-0001-9743-0442}}
\email{gustav.uhre.jakobsen@physik.hu-berlin.de}
\affiliation{Max Planck Institute for Gravitational Physics (Albert Einstein Institute), Am M\"uhlenberg 1, 14476 Potsdam, Germany}
\affiliation{%
  Institut f\"ur Physik und IRIS Adlershof, Humboldt Universit\"at zu Berlin,
  Zum Gro{\ss}en Windkanal 2, 12489 Berlin, Germany
}

\author{Gustav Mogull\,\orcidlink{0000-0003-3070-5717}}
\email{gustav.mogull@aei.mpg.de} 
\affiliation{Max Planck Institute for Gravitational Physics (Albert Einstein Institute), Am M\"uhlenberg 1, 14476 Potsdam, Germany}
\affiliation{%
  Institut f\"ur Physik und IRIS Adlershof, Humboldt Universit\"at zu Berlin,
  Zum Gro{\ss}en Windkanal 2, 12489 Berlin, Germany
}

\begin{abstract}
Effective-one-body (EOB) waveforms employed by the
    LIGO-Virgo-KAGRA Collaboration have primarily been developed by
    resumming the post-Newtonian expansion of the relativistic
    two-body problem. Given the recent significant advancements in
    post-Minkowskian (PM) theory and gravitational self-force
    formalism, there is considerable interest in creating waveform
    models that integrate information from various perturbative
    methods in innovative ways.  This becomes particularly crucial
    when tackling the accuracy challenge posed by upcoming
    ground-based detectors (such as the Einstein Telescope and Cosmic
    Explorer) and space-based detectors (such as LISA, TianQin or Taiji)
    expected to operate in the next decade. In this context, we
    present the derivation of the first spinning EOB Hamiltonian that
    incorporates PM results up to three-loop order: the SEOB-PM model.
   The model accounts for the complete hyperbolic motion,
    encompassing nonlocal-in-time tails. To evaluate its accuracy,
    we compare its predictions for the conservative scattering angle,
    augmented with dissipative contributions, against
    numerical-relativity data of non-spinning and spinning equal-mass
    black holes. We observe very good agreement, comparable, and in
    some cases slightly better to the recently proposed $w_{\rm
      EOB}$-potential model, of which the SEOB-PM model is a
    resummation around the probe limit. Indeed, in the probe 
limit, the SEOB-PM
    Hamiltonian and scattering angles reduce to the one of a test mass in Kerr
    spacetime. Once complemented with nonlocal-in-time contributions
    for bound orbits, the SEOB-PM Hamiltonian can be utilized to
    generate waveform models for spinning black holes on
    quasi-circular orbits.
\end{abstract}

\maketitle

\section{Introduction}

The observation of gravitational waves (GWs) from coalescing
  binary black holes (BHs) and neutron stars (NSs) provides a unique
  opportunity to probe fundamental physics, dynamical gravity and
  matter under extreme conditions~\cite{Abbott:2016blz,LIGOScientific:2017vwq,LIGOScientific:2018cki,LIGOScientific:2021sio,LIGOScientific:2021aug}. Having access to a large number of
  GW signals --- more than  100 published observations by the LIGO-Virgo-KAGRA (LVK) Collaboration and independent analyses
  \cite{KAGRA:2021vkt,Nitz:2021zwj,Wadekar:2023gea} at the time of writing
  --- permits us to shed light on the astrophysical scenarios
  responsible for the formation of these binary systems~\cite{KAGRA:2021duu}.  Successful
  GW searches, precise inference of astrophysical and cosmological properties, and correct
  identifications of sources require detailed knowledge of the
  expected signals. This is achieved employing waveform models that
  are built by combining the best available methods to solve the two-body problem in General Relativity (GR). 

On one side, for the
  inspiral stage of the binary coalescence, we can solve Einstein's
  equations analytically, but approximately, in (i) the weak-field and
  small-velocity limit (i.e., in post-Newtonian (PN) theory~\cite{Futamase:2007zz,
Blanchet:2013haa,Porto:2016pyg,Schafer:2018kuf,Levi:2018nxp}), (ii) in the weak-field 
regime (i.e., in post-Minkowskian (PM) theory~\cite{Westpfahl:1979gu,Westpfahl:1980mk,Bel:1981be,Westpfahl:1985tsl,schafer1986adm,Ledvinka:2008tk,Damour:2016gwp,Cheung:2018wkq,Bern:2019nnu,Buonanno:2022pgc}),
  and (iii) in the small mass-ratio limit 
  (i.e., in the gravitational-self force (GSF) formalism~\cite{Mino:1996nk,Quinn:1996am,Barack:2001gx,Barack:2002mh,Gralla:2008fg,Detweiler:2008ft,Keidl:2010pm,vandeMeent:2017bcc,Pound:2012nt,Pound:2019lzj,Gralla:2021qaf,Pound:2021qin,Warburton:2021kwk}). On the
  other side, for the late inspiral, merger and ringdown stages, we
  can solve the Einstein's equations numerically~\cite{Pretorius:2005gq,Campanelli:2005dd,Baker:2005vv} on supercomputers,
  obtaining highly accurate GW predictions. Performing simulations in numerical relativity (NR), 
however, is time consuming. Thus, NR cannot be used alone to build the several hundred
  thousands (millions) waveform models or templates that are used in
  matched-filtering searches (follow-up Bayesian analysis)~\cite{KAGRA:2021vkt}. Importantly, analytical 
and numerical results need to be combined synergistically to achieve the accuracy 
that is needed. This is obtained through the effective-one-body (EOB) formalism~\cite{Buonanno:1998gg,Buonanno:2000ef,Damour:2000we,Damour:2001tu,Buonanno:2005xu} that 
maps the two-body dynamics onto the dynamics of a test mass~\cite{Damour:2000we,Damour:2008qf,Damour:2014sva,Khalil:2020mmr,Khalil:2023kep} (or test spin~\cite{Barausse:2009xi,Barausse:2011ys,Vines:2016unv}) moving in a deformed 
Schwarzschild or Kerr spacetime, the deformation being the mass ratio. The EOB formalism 
also predicts the full coalescence waveform through physically motivated ansatze 
for the merger, and BH perturbation theory, and it can be made highly accurate through calibration to NR~\cite{Buonanno:2006ui,Buonanno:2007pf,Damour:2007yf,Pan:2011gk,Damour:2012ky,Pan:2013rra,Taracchini:2013rva,Bohe:2016gbl,Babak:2016tgq,Nagar:2018zoe,Ossokine:2020kjp,Akcay:2020qrj,Gamba:2021ydi,Pompili:2023tna,Ramos-Buades:2023ehm}.

So far, EOB waveform models employed by the LVK Collaboration have been built on resummations of the PN expansion, except for Refs.~\cite{Ramos-Buades:2023ehm,vandeMeent:2023ols}, which includes second-order GSF results~\cite{Warburton:2021kwk} for the gravitational modes and radiation-reaction force. Given the recent results in PM~\cite{Cheung:2018wkq,Bern:2019nnu,Bjerrum-Bohr:2019kec,Kalin:2019rwq,Kalin:2020fhe,Kalin:2020mvi,Cristofoli:2020uzm,Mogull:2020sak,Jakobsen:2021smu,Bern:2021yeh,Dlapa:2021vgp,Travaglini:2022uwo,Bjerrum-Bohr:2022blt,Kosower:2022yvp} and GSF~\cite{Pound:2021qin,Warburton:2021kwk} theories, there is now great interest 
in exploring and developing waveform models that assemble the information from all the different 
perturbative methods, and in novel ways. This is particularly important when addressing the accuracy challenge posed by 
ever more sensitive detectors operating in the next decade: on the ground, such as the Einstein Telescope~\cite{Punturo:2010zz} and Cosmic Explorer~\cite{Reitze:2019iox}, 
and in space, such as LISA~\cite{LISA:2017pwj}, TianQin or Taiji, which demand an improvement of the waveforms by two orders of magnitude or more, depending on the parameter space~\cite{Purrer:2019jcp,Kunert:2021hgm,LISAConsortiumWaveformWorkingGroup:2023arg}, and the inclusion 
of all physical effects (spin-precession, eccentricity, matter).
We remark that the LVK Collaboration also employs waveform models for the inspiral, merger and ringdown 
in frequency-domain by combining PN, EOB and NR results (i.e., the phenomenological templates~\cite{Ajith:2007qp,Pratten:2020ceb}), and 
waveforms that interpolate directly NR simulations (i.e., the NR-surrogate models~\cite{Blackman:2017pcm,Varma:2019csw}).

Building on previous work~\cite{Bini:2017xzy,Bini:2018ywr,Antonelli:2020ybz,Antonelli:2019ytb,Khalil:2022ylj,Khalil:2023kep},
in this paper we present the first EOB Hamiltonian for spinning bodies based on the PM expansion (henceforth, SEOB-PM) that includes complete non-spinning \cite{Bern:2021yeh,Dlapa:2021vgp,Dlapa:2022lmu,Dlapa:2023hsl,Damgaard:2023ttc} and spinning information \cite{Vines:2018gqi,Bern:2020buy,Kosmopoulos:2021zoq,Jakobsen:2021zvh,Guevara:2018wpp,Chen:2021kxt,Jakobsen:2022fcj,Jakobsen:2022zsx,FebresCordero:2022jts,Jakobsen:2023ndj,Jakobsen:2023hig} through 4PM order for hyperbolic motion, with additional corrections at 5PM. Our PM counting is a physical one for compact objects, such as black holes and neutron stars, spin orders contributing as well as loop orders. We  assess the accuracy of the SEOB-PM Hamiltonian by comparing its predictions for the scattering angle with available NR results~\cite{Damour:2014afa,Hopper:2022rwo,Rettegno:2023ghr} for nonspinning and 
spinning bodies with equal masses and equal spins. We contrast our 
model predictions with the non-spinning EOB model of Ref.~\cite{Khalil:2022ylj}, which is also based on an EOB Hamiltonian,  and the $w_{\rm EOB}$-model, which was developed in Refs.~\cite{Damour:2022ybd,Rettegno:2023ghr}. 
We also discuss its improvement against its PN counterpart and its comparison 
against the $w_{\rm EOB}$ model in the probe limit (and unequal-mass scatterings) since, differently from $w_{\rm EOB}$, the SEOB-PM model does reduce to the probe limit (i.e., it reduces to the Schwarzschild and Kerr results).

Importantly, the SEOB-PM model entails (or is based on) a Hamiltonian. Thus, it can be used to describe the two-body dynamics for bound orbits, and combined with a suitable radiation-reaction force and gravitational modes, it can be employed to generate waveform models for binary BHs with spins. To describe waveforms from quasi-circular orbits, Ref.~\cite{Buonanno:2024byg} has augmented such a Hamiltonian with nonlocal-in-time terms at 4PN order~\cite{Damour:2016abl,Marchand:2017pir,Foffa:2019rdf,Foffa:2019yfl,Blumlein:2020pog} for bound orbits.
The local-in-time contributions to the 4PM non-spinning Hamiltonian,
valid both for bound and unbound orbits, have also now been derived explicitly~\cite{Dlapa:2024cje},
although (for now) the bound-orbit tail contributions are only accessible via PN expansion for quasi-circular orbits.

The paper is organized as follows. After summarizing the notation used in this work, in Sec.~\ref{sec:pert} we review the perturbative conservative and dissipative contributions to the 
scattering angle, and also estimate the effect of the recoil on the scattering angles of BHs carrying different 
spin magnitudes. In Sec.~\ref{sec:probeAngle}, we highlight the scattering angle in the probe limit, i.e.~for a test-mass in a hyperbolic orbit 
about the Kerr spacetime. In Sec.~\ref{sec:EOB} we derive the SEOB-PM Hamiltonian,
resumming information at both 4PM and 5PM orders,
which is the main result of this paper.
In Sec.~\ref{sec:plots} we assess 
its validity by comparing its predictions for the scattering angle with NR data of Refs.~\cite{Hopper:2022rwo,Rettegno:2023ghr}, and 
also with the so-called $w_{\rm EOB}$-potential model proposed in those papers.
Finally, we summarize our main conclusions in Sec.~\ref{sec:concl}.

\subsection*{Notation}

Henceforth, we work in the (initial) center-of-mass (CoM) frame and employ natural units $G=c=1$.
For our spinning two-body system,
consisting of two scattered massive bodies (BHs),
we introduce the following combinations of masses $m_1$, $m_2$:
\begin{align}\label{eq:masses}
M&=m_1+m_2\,,\,\,
\nu=\frac{\mu}M=\frac{m_1m_2}{M^2}\,,\,\,
\delta=\frac{m_1-m_2}{M}\,,
\end{align}
so $\delta=\sqrt{1-4\nu}$ with  $m_1\ge m_2$.
We also introduce the total energy $E=E_1+E_2$ and effective energy $E_{\rm eff}$,
which are related by the energy map:

\begin{subequations}
\begin{align}
  \Gamma& \equiv \frac{E}{M}=\sqrt{1+2\nu(\gamma-1)}\,,\label{eq:energyMap}\\
  \gamma& \equiv \frac{E_{\rm eff}}{\mu}=\frac{E^2-m_1^2-m_2^2}{2m_1m_2}\,.\label{eq:Gamma}
\end{align}
\end{subequations}
The boost factor $\gamma>1$ is given by 
\begin{align}
  \gamma=\frac1{\sqrt{1-v^2}}\,,
\end{align}
where $v$ is the relativistic relative velocity.

\begin{figure}
  \centering
  \includegraphics[width=.4\textwidth]{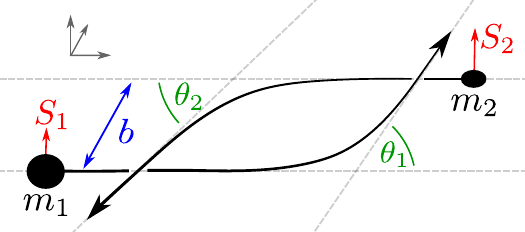}
  \caption{Kinematic setup of a planar two-body scattering event, with two separate scattering angles $\theta_1$, $\theta_2$
  in the dissipative case. 
  For the conservative dynamics, $\theta_1=\theta_2=\theta_{\rm cons}$.
  We specialize to aligned spins, with directed spin lengths $a_1=S_1/m_1$, $a_2=S_2/m_2$.
  (Diagram reproduced with minor edits from Ref.~\cite{Antonelli:2020aeb}, with permission of the authors.)}
  \label{fig:scattering}
\end{figure}

The initial CoM frame, as defined with respect to the \emph{incoming} momenta, is:
$p_1^\mu=(E_1,\vct{p}_\infty/\Gamma)$ and $p_2^\mu=(E_2,-\vct{p}_\infty/\Gamma)$,
where $p_i^2=m_i^2$ and $p_1\cdot p_2=\gamma\, m_1\,m_2$ --- see \Fig{fig:scattering}.
The relative CoM momentum at past infinity $\vct{p}_\infty$ has magnitude
\begin{align}
  p_\infty=|\vct{p}_\infty|=\mu\,v_\infty=\mu\sqrt{\gamma^2-1}\,.
\end{align}
The two bodies are initially separated by the impact parameter $\vct{b}$,
with $\vct{b}\cdot\vct{p}_\infty=0$ and $b=|\vct{b}|$.
As the motion evolves, we use the dynamical
relative position and momentum vectors $\vct{r}$, $\vct{p}$ to describe the two-body motion:
\begin{align}
  \vct{p}^2&=p_r^2+\frac{L^2}{r^2}\,, &
  p_r&=\hat{\vct{r}}\cdot\vct{p}\,, &
  \vct{L}=\vct{r}\cross\vct{p}\,,
\end{align}
where $r=|\vct{r}|$ and $\hat{\vct{r}}=\vct{r}/r$;
$\vct{L}$ is the canonical orbital angular momentum with directed length $L=b\,\pin /\Gamma$.
In the limit $r\to\infty$, we have $\vct{p}\to\vct{p}_\infty$. 
We also introduce $u=M/r$, which we use as the PM counting parameter,
and $\ell=L/(M\mu)$ is the dimensionless orbital angular momentum.

Finally, we specialize to spin vectors $\vct{S}_i$ of the two bodies aligned
with the orbital angular momentum $\vct{L}$.
The total angular momentum $J$ is then given by
\begin{align}\label{eq:angMom}
  J&=L+S_1+S_2\,, &
  S_i&=m_ia_i=m_i^2\chi_i\,.
\end{align}
Here $a_i$ are the directed spin lengths ---
for Kerr BHs, these are the radii of the ring singularities,
and the dimensionless spin lengths are $-1<\chi_i<1$.
Including, for clarity, units one has $S_i=G m_i^2 \chi_i/c$.
We also introduce the combinations
\begin{align}
  a_\pm&=M\chi_\pm=a_1\pm a_2\,.
\end{align}
Results from the PM-scattering literature
are often given in terms of the \emph{covariant} orbital angular momentum $L_{\rm cov}$.
For aligned spins, the covariant orbital angular momentum $L_{\rm cov}$ is related to
the total angular momentum $J$ by:
\begin{align}
  J&=L_{\rm cov}+E_1a_1+E_2a_2\,.
\end{align}
Using \Eqn{eq:angMom} we may eliminate the total angular momentum $J$,
and learn that
\begin{align}
  L_{\rm cov}=L-\frac{E-M}{2}\left(a_+-\frac{\delta}{\Gamma}a_-\right)\,.
\end{align}
We use this to re-express $L_{\rm cov}$ in terms of $L$.

\section{
  Perturbative Relativistic Scattering
}
\label{sec:pert}

Let us focus on two-body relativistic BH scattering events --- depicted in \Fig{fig:scattering}.
In order to prepare the ground for the scattering angles derived with the spinning EOB model based on PM (SEOB-PM) in Sec.~\ref{sec:EOB}, we focus first on the 
conservative and dissipative PM dynamics in Secs.~\ref{sec:consAngle} and~\ref{sec:dissAngle}, respectively, and then consider 
the probe motion in Sec.~\ref{sec:probeAngle}.
In connection with the PM regime, we also consider the PN and GSF expansions.
Each of the three perturbative regimes may be defined by assuming certain combinations of the initial data to be small, of order $\eps\ll1$.
Using $M$ as a scale, the three dimensionless parameters,
$1/\ell$, $v$ and $\nu$, together with the spins $S_i$, fully describe the initial state.
The scalings of these parameters together with $u$ for the different perturbative schemes are summarized in \Tab{table:scalings}.
Notice that, in this \emph{physical} counting scheme,
the PM expansion does not align with the loop expansion, as it does in a \emph{formal} counting ---
on dimensional grounds, powers of the spins $S_i=Gm_i^2\chi_i/c$ come with additional factors of $v$, $\ell$ and/or $u$, which are included in the counting.
The scalings of other variables may be inferred by expressing them in terms of the basic ones in \Tab{table:scalings}.

\begin{table}
  \centering
  \begin{tabular}{|m{3em}|m{3em}|m{3em}|m{3em}|m{3em}|}
    \hhline{~|-|-|-|-|}
    \multicolumn{1}{c|}{} &
    \cellcolor{lightgray}  \hfil $\ell^{-1}$ &
    \cellcolor{lightgray} \hfil $v^2$ &
    \cellcolor{lightgray} \hfil $\nu$ &
    \cellcolor{lightgray} \hfil $u$
    \\ 
    \hline
    \cellcolor{lightgray} PM &
    \hfil $\sim\eps$ &
    \hfil $\sim1$ &
    \hfil $\sim 1$ &
    \hfil $\sim\eps$
    \\ 
    \hline
    \cellcolor{lightgray} PN &
    \hfil $\sim\sqrt{\eps}$ &
    \hfil $\sim\eps$ &
    \hfil $\sim 1$ &
    \hfil $\sim\eps$
    \\ 
    \hline
    \cellcolor{lightgray} GSF &
    \hfil $\sim1$ &
    \hfil $\sim1$ &
    \hfil $\sim\eps$ &
    \hfil $\sim1$
    \\
    \hline
  \end{tabular}
  \caption{Scalings of the dimensionless variables $\ell$, $v$, $\nu$ and $u$ in the PM, PN and GSF regimes,
  with $\eps\ll 1$ a counting parameter.
  The dimensionless spins are taken with
  scaling $\chi_i\sim1$ in all three regimes.
  }
\label{table:scalings}
\end{table} 

\subsection{
  Conservative Scattering Angle
}
\label{sec:consAngle}

Let us first consider the ideal setting of conservative scattering,
whose main characteristic is that the total energy and CoM angular momentum, $E$ and $L$, are conserved (implying in turn also that, e.g., $v$, $b$ and $p_\infty$ are conserved).
In this case the motion is completely symmetric and fully described by the scattering angle which we label: $\theta_{\rm cons}$.
This angle is related to the conservative momentum impulse $\Del p^\mu_{\rm cons}=\Del p^\mu_{1,\rm cons}=-\Del p^\mu_{2,\rm cons}$ via
\begin{align}\label{eq:conservativeAngle}
   \sin\!
   \Big(
    \frac{\theta_{\rm cons}}{2}
    \Big)
   =
  \Gamma\,\frac{
    |\Del p_{\rm cons}^\mu|
   }{
    2 p_{\infty}
   }
   \ ,
\end{align}
which may be derived by geometrical arguments. In this setting the two angles $\theta_i$ of Fig.~\ref{fig:scattering} are equal and denoted by $\theta_{\rm cons}$.

In this conservative approximation, the dynamics and scattering angle may equally well be described by a Hamiltonian.
Rather than the Hamiltonian, in this section we find it useful to use the effective potential $w$ defined from the mass-shell constraint:
\begin{align}\label{eq:impetusTrue}
  p_r^2
  &=
  \pin^2
  -
  \frac{L^2}{r^2}
  +w(E_{\rm eff},L,r;a_\pm)
  \,.
\end{align}
Using the Hamilton-Jacobi formalism (see e.g.~\Rcite{Damour:2017zjx}) the relationship between the potential and the angle is:
\begin{align}\label{eq:angleFromPR}
  \theta+\pi
      =
      -2\int_{r_{\rm min}}^\infty\!{\rm d}r\frac{\partial}{\partial L}
      p_{r}
      \,,
\end{align}
where $r_{\rm min}$ is the closest point of approach defined as the largest root of $p_{r}(r_{\rm min})=0$.
Here, we have omitted the `cons' subscript on the angle, because in Sec.~\ref{sec:dissAngle} we also define a ``dissipative'' effective potential.
This formula is generic, and applies both to the conservative setting
discussed here and the dissipative effects to be discussed in Sec.~\ref{sec:dissAngle}.

By way of \Eqn{eq:angleFromPR}, the scattering angle $\theta$ and the effective potential $w$ are in one-to-one correspondence. An expression for the potential in isotropic gauge in terms of the angle is
given by the Firsov formula discussed in Refs.~\cite{Kalin:2019rwq,Damour:2022ybd}.
While the angle is gauge-invariant, the potential is not;
thus, it is uniquely determined by the angle only when a gauge condition is imposed. Nevertheless, the potential has certain advantages over the angle:
it is finite in the PN limit and it has a simple expression (compared with the angle, see e.g.~Ref~\cite{Kol:2021jjc}) in the probe limit. 
The angle generally depends on the dimensionless initial state variables ($\gamma$, $\nu$, $\ell$ and $\chi_i$), while the potential, in addition to these, also depends on the relative position, $u$ (and on $M$ to balance dimensions).

The PM expansions of the angle and potential are:
\begin{subequations}
\begin{align}
  \theta
  &=
  \sum_{n\geq1} 
  \frac{\theta^{(n)}}{\ell^n}\,,\label{eq:angle}
  \\
  w
  &=
  \sum_{n\geq1} u^n
  w^{(n)}\,,
  \label{eq:wPMExpansion}
\end{align}
\end{subequations}
$n$ counting the PM orders with PM expansion variables $1/\ell$ and $u$ (see \Tab{table:scalings}).
We define also $m$PM accurate scattering angles: $\theta^{m{\rm PM}}=\sum_{n=1}^m \theta^{{(n)}}/\ell^n$.
We may further expand the PM coefficients with respect to the BHs' spins:
\begin{subequations}
\begin{align}
  \label{eq:angleSpinExpansion}
  \theta^{(n)}
  &=
  \sum_{s=0}^{n-1}\sum_{i=0}^s
      \theta^{(n)}_{(s-i,i)}\delta^{\sig(i)}\chi_+^{s-i}\chi_-^{i},
      \\
  w^{(n)}
  &=
  \sum_{s=0}^{n}(\ell u)^{\sig(s)}\sum_{i=0}^s
    w^{(n)}_{(s-i,i)}
    \delta^{\sig(i)}\chi_+^{s-i}\chi_-^{i}.
      \label{eq:spinExpansion}
\end{align}
\end{subequations}
Here, $s$ counts the spin orders and $i$ the powers of $\chi_-$.
The function $\sigma$ is given as
\begin{align}
  \sig(n)
  =
  \begin{cases}
        0\,, & n \text{ even}, \\
        1\,, & n \text{ odd},
  \end{cases}
\end{align}
and controls the introduction of $\delta$ to terms with odd powers of $\chi_-$ and a power of $\ell \, u$ for odd $s$ in the potential.

\begin{table}
  \centering
  \begin{tabular}{|m{5em}|m{2.9em}|m{2.9em}|m{2.9em}|m{2.9em}|m{2.9em}|m{2.9em}|}
    \hhline{~|-|-|-|-|-|-|}
    \multicolumn{1}{c|}{} & \cellcolor{lightgray} \hfil $S^0$ & \cellcolor{lightgray} \hfil $S^1$ & \cellcolor{lightgray} \hfil $S^2$ & \cellcolor{lightgray} \hfil $S^3$ & \cellcolor{lightgray} \hfil $S^4$ & \cellcolor{lightgray} \hfil $S^5$ \\ 
    \hline
    \cellcolor{lightgray} tree level & \cellcolor{Cerulean} \hfil 1PM & \cellcolor{Green} \hfil 2PM & \cellcolor{YellowGreen} \hfil 3PM & \cellcolor{Goldenrod} \hfil 4PM & \cellcolor{BurntOrange} \hfil 5PM & \cellcolor{Red} \hfil 6PM \\ 
    \hline
    \cellcolor{lightgray} 1-loop & \cellcolor{Green} \hfil 2PM & \cellcolor{YellowGreen} \hfil 3PM & \cellcolor{Goldenrod} \hfil 4PM & \cellcolor{BurntOrange} \hfil 5PM & \cellcolor{Red} \hfil 6PM & \hfil 7PM \\ 
    \hline
    \cellcolor{lightgray} 2-loop & \cellcolor{YellowGreen} \hfil 3PM & \cellcolor{Goldenrod} \hfil 4PM & \cellcolor{BurntOrange} \hfil 5PM & \hfil 6PM & \hfil 7PM & \hfil 8PM \\
    \hline
    \cellcolor{lightgray} 3-loop & \cellcolor{Goldenrod} \hfil 4PM & \cellcolor{BurntOrange} \hfil 5PM & \hfil 6PM & \hfil 7PM & \hfil 8PM & \hfil 9PM \\
    \hline
    \cellcolor{lightgray} 4-loop & \hfil 5PM & \hfil 6PM & \hfil 7PM & \hfil 8PM & \hfil 9PM & \hfil 10PM \\
    \hline
  \end{tabular}
\caption{
  The relation between loop orders, spin orders and physical PM orders (as adopted in this work), with currently known results for the aligned-spin scattering angle $\theta$ colored in.
The tree-level all-order-spin angle is given in \Eqn{eq:treeAngle}.
The 1-loop angle is known up to 
quartic $G^6$ order in spin~\cite{Guevara:2018wpp,Chen:2021kxt},
with higher-order predictions in Refs.~\cite{Bern:2022kto,Aoude:2023vdk,Bautista:2023szu}.
The 2-loop $G^3$ conservative dynamics were derived in Refs.~\cite{Bern:2019nnu,Bern:2019crd,Cheung:2020gyp,DiVecchia:2020ymx,Kalin:2020fhe},
extended to include radiation~\cite{Damour:2020tta,Herrmann:2021lqe,Herrmann:2021tct,DiVecchia:2021bdo,Heissenberg:2021tzo,Bjerrum-Bohr:2021din,Damgaard:2021ipf,Jakobsen:2022psy,Kalin:2022hph}
and $G^4$, $G^5$ linear- and quadratic-in-spin effects~\cite{Jakobsen:2022fcj,Jakobsen:2022zsx,FebresCordero:2022jts}.
The 3-loop non-spinning $G^4$ terms were obtained in Refs.~\cite{Bern:2021yeh,Dlapa:2021vgp} (conservative)
and \cite{Dlapa:2022lmu,Dlapa:2023hsl,Damgaard:2023ttc} (full dissipative).
Spin-orbit effects ($G^5$) were incorporated in Refs.~\cite{Jakobsen:2023ndj,Jakobsen:2023hig}.
Finally, 5PM non-spinning conservative effects at 1GSF have most recently been derived in Ref.~\cite{Driesse:2024xad}.
}
\label{tab:PMtable}
\end{table}
\begin{table*}
  \centering
  \subfloat{
  \begin{tabular}{
    |
    m{4em}||
    m{4em}|
    m{4em}|
    m{4em}|
    m{4em}|
    m{4em}|
    }
    \hline
    \cellcolor{lightgray}
    \hfil$\theta$
    &\cellcolor{lightgray}
    \hfil
    0PN
    &\cellcolor{lightgray}
    \hfil
    1PN
    &\cellcolor{lightgray}
    \hfil
    2PN
    &\cellcolor{lightgray}
    \hfil
    3PN
    &\cellcolor{lightgray}
    \hfil
    4PN
    \\
    \hline
    \hline
    \cellcolor{lightgray}
    \hfil
    1PM $S^0$
    &
    ${\ell^{-1}v^{-1}}^{\vphantom{L}}$
    &
    ${\ell^{-1}v}^{\vphantom{L}}$
    &
    ${\ell^{-1}v^{3}}^{\vphantom{L}}$
    &
    ${\ell^{-1}v^{5}}^{\vphantom{L}}$
    &
    ${\ell^{-1}v^{7}}^{\vphantom{L}}$
    \\
    \hline
    \cellcolor{lightgray}
    \hfil
    2PM $S^0$
    &
    $\ $
    &
    ${\ell^{-2}}^{\vphantom{L}}$
    &
    ${\ell^{-2}v^2}^{\vphantom{L}}$
    &
    ${\ell^{-2}v^4}^{\vphantom{L}}$
    &
    ${\ell^{-2}v^{6}}^{\vphantom{L}}$
    \\
    \hline
    \cellcolor{lightgray}
    \hfil
    3PM $S^0$
    &
    ${\ell^{-3}v^{-3}}^{\vphantom{L}}$
    &
    ${\ell^{-3}v^{-1}}^{\vphantom{L}}$
    &
    \cellcolor{YellowGreen}
    ${\ell^{-3}v}^{\vphantom{L}}$
    &
    \cellcolor{YellowGreen}
    ${\ell^{-3}v^3}^{\vphantom{L}}$
    &
    \cellcolor{YellowGreen}
    ${\ell^{-3}v^5}^{\vphantom{L}}$
    \\
    \hline
    \cellcolor{lightgray}
    \hfil
    4PM $S^0$
    &
    $\ $
    &
    $\ $
    &
    \cellcolor{YellowGreen}
    ${\ell^{-4}}^{\vphantom{L}}$
    &
    \cellcolor{YellowGreen}
    ${\ell^{-4}v^{2}}^{\vphantom{L}}$
    &
    \cellcolor{YellowGreen}
    ${\ell^{-4}v^4(*)}^{\vphantom{L}}$
    \\
    \hline
    \cellcolor{lightgray}
    \hfil
    5PM $S^0$
    &
    \hfil
    $\ell^{-5}v^{-5}$
    &
    $\ell^{-5}v^{-3}$
    &
    \cellcolor{YellowGreen}
    ${\ell^{-5}v^{-1}}$
    &
    \cellcolor{YellowGreen}
    ${\ell^{-5}v}$
    &
    \cellcolor{YellowGreen}
    ${\ell^{-5}v^3(*)}$
    \\
    \hline
    \hline
    \cellcolor{lightgray}
    \hfil
    3PM $S^2$
    &
    $\ $
    &
    $\ $
    &
    ${\ell^{-3}v}^{\vphantom{L}}$
    &
    ${\ell^{-3}v^3}^{\vphantom{L}}$
    &
    ${\ell^{-3}v^5}^{\vphantom{L}}$
    \\
    \hline
    \cellcolor{lightgray}
    \hfil
    4PM $S^2$
    &
    $\ $
    &
    $\ $
    &
    ${\ell^{-4}}^{\vphantom{L}}$
    &
    ${\ell^{-4}v^{2}}^{\vphantom{L}}$
    &
    ${\ell^{-4}v^4 }^{\vphantom{L}}$
    \\
    \hline
    \cellcolor{lightgray}
    \hfil
    5PM $S^2$
    &
    $\ $
    &
    $\ $
    &
    $ \ell^{-5} v^{-\!1}$
    &
    ${ \ell^{-5}v}^{\vphantom{L}}$
    &
    \cellcolor{YellowGreen}
    ${ \ell^{-5}v^3 }^{\vphantom{L}}$
    \\
    \hline
    \hline
    \cellcolor{lightgray}
    \hfil
    5PM $S^4$
    &
    $\ $
    &
    $\ $
    &
    $\ $
    &
    $\ $
    &
    $\ell^{-5}v^3$
    \\
    \hline
    \noalign{\vskip 2mm}
    \hline 
    \cellcolor{lightgray}
    \hfil $\theta$
    &
    \cellcolor{lightgray}
    \hfil
    0.5PN
    &
    \cellcolor{lightgray}
    \hfil
    1.5PN
    &
    \cellcolor{lightgray}
    \hfil
    2.5PN
    &
    \cellcolor{lightgray}
    \hfil
    3.5PN
    &
    \cellcolor{lightgray}
    \hfil
    4.5PN
    \\
    \hhline{=||=|=|=|=|=|}
    \cellcolor{lightgray}
    \hfil
    2PM $S^1$
    &
    $\ $
    &
    ${\ell^{-2}v}^{\vphantom{L}}$
    &
    ${\ell^{-2}v^3}^{\vphantom{L}}$
    &
    ${\ell^{-2}v^{5}}^{\vphantom{L}}$
    &
    ${\ell^{-2}v^{7}}^{\vphantom{L}}$
    \\
    \hline
    \cellcolor{lightgray}
    \hfil
    3PM $S^1$
    &
    $\ $
    &
    ${\ell^{-3}}^{\vphantom{L}}$
    &
    ${\ell^{-3}v^{2}}^{\vphantom{L}}$
    &
    ${\ell^{-3}v^4}^{\vphantom{L}}$
    &
    ${\ell^{-3}v^6}^{\vphantom{L}}$
    \\
    \hline
    \cellcolor{lightgray}
    \hfil
    4PM $S^1$
    &
    $\ $
    &
    ${\ell^{-4}v^{-1}}^{\vphantom{L}}$
    &
    ${\ell^{-4}v}^{\vphantom{L}}$
    &
    \cellcolor{YellowGreen}
    ${\ell^{-4}v^{3}}^{\vphantom{L}}$
    &
    \cellcolor{YellowGreen}
    ${\ell^{-4}v^5}^{\vphantom{L}}$
    \\
    \hline
    \cellcolor{lightgray}
    \hfil
    5PM $S^1$
    &
    $\ $
    &
    $\ $
    &
    ${\ell^{-5}}^{\vphantom{L}}$
    &
    \cellcolor{YellowGreen}
    ${\ell^{-5}v^{2}}^{\vphantom{L}}$
    &
    \cellcolor{YellowGreen}
    ${\ell^{-5}v^4}^{\vphantom{L}}$
    \\
    \hline
    \hline
    \cellcolor{lightgray}
    \hfil
    4PM $S^3$
    &
    $\ $
    &
    $\ $
    &
    $\ $
    &
    ${\ell^{-4}v^{3}}^{\vphantom{L}}$
    &
    ${\ell^{-4}v^5 }^{\vphantom{L}}$
    \\
    \hline
    \cellcolor{lightgray}
    \hfil
    5PM $S^3$
    &
    $\ $
    &
    $\ $
    &
    $\ $
    &
    ${\ell^{-5}v^{2}}^{\vphantom{L}}$
    &
    ${\ell^{-5}v^4}^{\vphantom{L}}$
    \\
    \hline
\end{tabular}
  }
  \subfloat{
    \begin{tabular}{
      |
      m{4em}||
      m{4em}|
      m{4em}|
      m{4em}|
      m{4em}|
      m{4em}|
      }
      \hline
      \cellcolor{lightgray}
      \hfil $w$
      &
      \cellcolor{lightgray}
      \hfil
      0PN
      &
      \cellcolor{lightgray}
      \hfil
      1PN
      &
      \cellcolor{lightgray}
      \hfil
      2PN
      &
      \cellcolor{lightgray}
      \hfil
      3PN
      &
      \cellcolor{lightgray}
      \hfil
      4PN
      \\
      \hline
      \hline
      \cellcolor{lightgray}
      \hfil
      1PM $S^0$
      &
      ${u}^{\vphantom{L}}$
      &
      ${u\, v^2}^{\vphantom{L}}$
      &
      ${u\, v^{4}}^{\vphantom{L}}$
      &
      ${u\, v^{6}}^{\vphantom{L}}$
      &
      $u\, v^8$
      \\
      \hline
      \cellcolor{lightgray}
      \hfil
      2PM $S^0$
      &
      $\ $
      &
      ${u^2 }^{\vphantom{L}}$
      &
      ${u^2 v^2}^{\vphantom{L}}$
      &
      ${u^2 v^{4}}^{\vphantom{L}}$
      &
      $u^2 v^6$
      \\
      \hline
      \cellcolor{lightgray}
      \hfil
      3PM $S^0$
      &
      $\ $
      &
      $\ $
      &
      \cellcolor{YellowGreen}
      ${u^3 }^{\vphantom{L}}$
      &
      \cellcolor{YellowGreen}
      ${u^3 v^2}^{\vphantom{L}}$
      &
      \cellcolor{YellowGreen}
      $u^3 v^4$
      \\
      \hline
      \cellcolor{lightgray}
      \hfil
      4PM $S^0$
      &
      $\ $
      &
      $\ $
      &
      \cellcolor{YellowGreen}
      $\ $
      &
      \cellcolor{YellowGreen}
      ${u^4}^{\vphantom{L}}$
      &
      \cellcolor{YellowGreen}
      $u^4 v^2 (*)$
      \\
      \hline
      \cellcolor{lightgray}
      \hfil
      5PM $S^0$
      &
      $\ $
      &
      $\ $
      &
      \cellcolor{YellowGreen}
      $\ $
      &
      \cellcolor{YellowGreen}
      $\ $
      &
      \cellcolor{YellowGreen}
      $u^5 (*)$
      \\
      \hline
      \hline
      \cellcolor{lightgray}
      \hfil
      3PM $S^2$
      &
      $\ $
      &
      $\ $
      &
      $ u^3 $
      &
      $ u^3 v^2 $
      &
      $ u^3 v^4 $
      \\
      \hline
      \cellcolor{lightgray}
      \hfil
      4PM $S^2$
      &
      $\ $
      &
      $\ $
      &
      $\ $
      &
      $ u^4 $
      &
      $ u^4 v^2 $
      \\
      \hline
      \cellcolor{lightgray}
      \hfil
      5PM $S^2$
      &
      $\ $
      &
      $\ $
      &
      $\ $
      &
      $\ $
      &\cellcolor{YellowGreen}
      $ u^5 $
      \\
      \hline
      \hline
      \cellcolor{lightgray}
      \hfil
      5PM $S^4$
      &
      $\ $
      &
      $\ $
      &
      $\ $
      &
      $\ $
      &
      $u^5$
      \\
      \hline
      \noalign{\vskip 2mm}
      \hline 
      \cellcolor{lightgray}
      \hfil $w$
      &
      \cellcolor{lightgray}
      \hfil
      0.5PN
      &
      \cellcolor{lightgray}
      \hfil
      1.5PN
      &
      \cellcolor{lightgray}
      \hfil
      2.5PN
      &
      \cellcolor{lightgray}
      \hfil
      3.5PN
      &
      \cellcolor{lightgray}
      \hfil
      4.5PN
      \\
      \hhline{=||=|=|=|=|=|}
      \cellcolor{lightgray}
      \hfil
      2PM $S^1$
      &
      $\ $
      &
      $ \ell u^3 $
      &
      $ \ell u^3 v^2 $
      &
      $ \ell u^3 v^4 $
      &
      $ \ell u^3 v^6 $
      \\
      \hline
      \cellcolor{lightgray}
      \hfil
      3PM $S^1$
      &
      $\ $
      &
      $\ $
      &
      $ \ell u^4 $
      &
      $ \ell u^4 v^2 $
      &
      $ \ell u^4 v^4 $
      \\
      \hline
      \cellcolor{lightgray}
      \hfil
      4PM $S^1$
      &
      $\ $
      &
      $\ $
      &
      $\ $
      &
      \cellcolor{YellowGreen}
      $ \ell u^5 $
      &
      \cellcolor{YellowGreen}
      $ \ell u^5 v^2 $
      \\
      \hline
      \cellcolor{lightgray}
      \hfil
      5PM $S^1$
      &
      $\ $
      &
      $\ $
      &
      $\ $
      &
      \cellcolor{YellowGreen}
      $\ $
      &
      \cellcolor{YellowGreen}
      $ \ell u^6 $
      \\
      \hline
      \hline
      \cellcolor{lightgray}
      \hfil
      4PM $S^3$
      &
      $\ $
      &
      $\ $
      &
      $\ $
      &
      $ \ell u^5 $
      &
      $ \ell u^5 v^2 $
      \\
      \hline
      \cellcolor{lightgray}
      \hfil
      5PM $S^3$
      &
      $\ $
      &
      $\ $
      &
      $\ $
      &
      $\ $
      &
      $ \ell u^6 $
      \\
      \hline
    \end{tabular}
  }
\caption{Schematic overview of the terms appearing in a combined PM and PN expansion of the angle (left) and potential (right).
  In these expansions, terms with the above scalings in $\ell$, $v$ and $u$ appear with
  coefficients depending only on the symmetric mass ratio $\nu$ and the dimensionless spins $\chi_\pm$
  (neither of which scales in the PM or PN expansions);
  empty cells are absent from the expansion.
  The explicitly shown terms are valid for the conservative dynamics,
  while the green shaded cells also pick up a 0.5PN subleading dissipative correction.
  The cells marked with a $(*)$ also have a tail contribution, implying $\log(v)$-dependence.
  The PN counting of the potential $w$ matches that of the Hamiltonian --- see, e.g.,~Ref.~\cite{Vines:2016qwa}.
  The analytic results of all PM rows (to all PN orders) are known except for 5PM $S^0$ ---
  see Table~\ref{tab:PMtable}. The green shading and tail terms of the 5PM $S^0$ row are thus only an expectation.
}
\label{tab:PN}
\end{table*}

As seen from \Eqn{eq:angleSpinExpansion}, the $n$PM angle gets contributions only from spin orders $s<n$.
Our physical PM counting, valid for compact objects, is different from the \textit{formal} PM counting often used in the PM literature,
which aligns with the loop order.
The relation between the physical PM counting relevant for spinning BHs and the formal loop-order PM counting is summarized in \Tab{tab:PMtable}.
Essentially, an $l$-loop result at spin order $s$ contributes to the $(l+s+1)$-PM order, so that both loops and spin orders climb up in PM orders.
Colored entries of \Tab{tab:PMtable} indicate known results.
The expansion coefficients of the angle depend non-trivially only on $\gamma$ and, when suitable variables are chosen and up to overall factors, on polynomials of the mass ratio (i.e., the mass polynomiality first observed in Refs.~\cite{Vines:2018gqi,Damour:2019lcq}).

The PM expansion of the potential $w$ takes a particularly simple form in
quasi-isotropic gauge, which is defined by requiring that the 
expansion coefficients of \Eqn{eq:spinExpansion} $w_{(j,i)}^{(n)}$ 
depend only on $\gamma$ and the masses, and that the sum on $s$
terminates at $n-1$ (i.e.~$w^{(n)}_{(n-i,i)}=0$ for all $i$). 
We refer to this potential generally as $w_{{\rm PM}}$;
or, up to a specified PM order, $w_{n {\rm PM}}$:
\begin{align}\label{eq:wEOBModel}
  w_{n{\rm PM}}
  =
  \sum_{m=1}^n
  \sum_{s=0}^{m-1}
  \ell^{\sig(s)}
  \sum_{i=0}^s
  w_{{\rm PM},(s-i,i)}^{(m)}
  \delta^{\sig(i)}\chi_+^{s-i}\chi_-^{i}
  \,.
\end{align}
In Refs.~\cite{Damour:2022ybd,Rettegno:2023ghr}\footnote{\label{counting} Note,
    however, that their inclusion of spin effects does not follow the
    physical PM counting introduced here.  Instead, their counting
    follows the formal PM counting.  Thus, their $n$PM model
    includes the colored entries of the first $n$ rows of
    Table.~\ref{tab:PMtable}, omitting the tree level $S^5$
    and the 3-loop $S^1$ contributions.}
it was shown that this potential defines a useful resummation of the angle:
simply by inserting $w_{n\text{PM}}$ into Eq.~\eqref{eq:angleFromPR},
one already finds a good agreement with the NR scattering angles (in particular, when incorporating dissipative effects as discussed in Sec.~\ref{sec:dissAngle}).  In the same two papers,
this computation of the scattering 
angles was dubbed $w_{\rm EOB}$.  In the present work, however, we prefer to
refer to a model as an EOB model only if it
reproduces the probe motion in the limit $\nu\to0$.  That is, from a PM
perspective, we require that the EOB model incorporates all PM orders in
the $\nu\to0$ limit, which is not the case for $w_{n\text{PM}}$.
The computation of the PM-expanded scattering angle through Eq.~\eqref{eq:wEOBModel} also coincides 
with Refs.~\cite{Kalin:2019rwq,Kalin:2019inp}, where it
was dubbed ``$f_{n}$-theory'' (for non-spinning 2PM dynamics).

The PM expansion of the potential $w$ in the SEOB-PM model (see Sec.~\ref{sec:EOB})
takes a more general form with dependence of $w_{(j,i)}^{(n)}$ on $ \ell u$ and non-zero $w_{(n-i,i)}^{(n)}$. 
Note, however, that the SEOB-PM $\nu$-deformation parameters, have the same simplicity as the $w_{\text{PM}}$ expansion coefficients.

Let us finally analyze the PN structure of the scattering angle and $w_{\rm PM}$ potential, which is schematically shown in Table~\ref{tab:PN}.
The PM/PN structure of $w_{\rm PM}$ is identical to the PM/PN structure of the SEOB-PM deformations to be introduced below.
Thus, we refer to it generically 
in Table~\ref{tab:PN} as $w$. In \Tab{tab:PN}, all scalings of $v$, $\ell$ and $u$
are shown in a combined PM and PN expansion up to 5PM and 4.5PN, respectively.
This table illustrates a key advantage of the potential over the angle, namely that in the angle, 0PN terms appear with arbitrarily high PM orders.
Instead, for the potential 0PN is completely determined by 1PM and generally, $n$PN is completely determined by $(n+1)$PM.
\Tab{tab:PN} also illustrates how spin pushes results to both higher PM and PN orders.
The green shaded cells indicate terms that receive a dissipative correction at a subleading 0.5PN order, 
which we discuss in the next section.
As stated above, full analytic PM results exist for all rows of \Tab{tab:PN} except the spinless row at 5PM.
For the non-spinning, conservative PN angle, PN resummations along the columns to 3PN order can be found in Ref.~\cite{Bini:2017wfr} (including, also, partial 4PN order results).

To illustrate these points, the 1PM row and 0PN columns are, respectively, given by the following two expressions:
\begin{subequations}
\begin{align}
  \label{eq:angle1PMand0PN}
  \theta
  &=
  \frac{2}{\ell v}
  \frac{
    1+v^2
  }{
    \sqrt{1-v^2}
  }
  +\mO \bigg(\frac{1}{\ell^2} \bigg)
  \\
  \label{eq:anglePN}
  &=
  2
  \text{arctan}
  \bigg(
  \frac{1}{\ell v}
  \bigg)
  +
  \mO(v^2)
  \ .
\end{align}
\end{subequations}
The expansion of the second line in $1/\ell$ clearly illustrates that the 0PN angle gets contributions at arbitrarily high (odd) PM orders. Yet both of these limits are included in the $w_{\rm 1PM}$ angle:
  \begin{align}\label{eq:weob1PM}
\theta^{w}_{\rm 1PM}
  &=
  2 \text{arctan}
  \Bigg(
    \frac{1}{\ell v}
    \frac{1+v^2}{\sqrt{1-v^2}}
  \Bigg)\,,
\end{align}
which clearly reproduces Eqs.~\eqref{eq:angle1PMand0PN} and~\eqref{eq:anglePN} in their respective limits.
This angle agrees with expressions found in Refs.~\cite{Kalin:2019rwq,Damour:2022ybd}.

\subsection{Dissipative Effects}
\label{sec:dissAngle}

Naturally, the true relativistic two-body system is dissipative.
As seen from the initial CoM frame, each BH loses energy,
and so the two momentum impulses $\Del p_i^\mu$ generally differ:
\begin{align}
  \Del p_1^\mu+\Del p_2^\mu
  =
  \Del P^\mu
  \neq 0
  \ .
\end{align}
The four vector $\Del P^\mu$ describes the total loss of linear momentum of the system.
If this vector has a non-zero spatial part with respect to the initial CoM frame, then the final CoM frame 
(defined from $p_i^\mu+\Delta p_i^\mu$) is different from the initial one.
This effect is referred to as recoil, and we define the spatial part of $\Del P^\mu=(\Del E,\Del \vct{P})$ with respect to the initial CoM frame as the recoil vector.\footnote{Another source of dissipation is horizon absorption~\cite{Poisson:1994yf,Saketh:2022xjb}, which we will not consider in this work.
}

A non-zero recoil implies, and is implied by, unequal scattering angles of the two bodies ---
see Fig.~\ref{fig:scattering}.
In this more general setting, the scattering of either body is computed by the geometric formula,
\begin{align}
  \cos \theta_i
  =
  \frac{
    (\vct{p}_i
  +
  \Del\vct{p_i})
  \cdot
  \vct{p}_i
  }{
    |\vct{p}_i
    +
    \Del\vct{p_i}|
    |\vct{p}_i|
  }
  \ .
\end{align}
With zero recoil, $\Del \vct{p}_1=-\Del \vct{p}_2$, and
the two angles are equal (using, also, $\vct{p}_1=-\vct{p}_2$).
At leading order in the PM expansion, one finds for the angle difference $\theta_1-\theta_2$ that
\begin{align}\label{eq:angleDifferenceAnalytic}
  \theta_1-\theta_2
  =
  \frac{\Gamma}{p_\infty}
  \bigg(
    \frac{\vct{\Delta P}\cdot 
  \vct{b}}{
    b
  }
  -
  \frac{\vct{\Delta P}
  \cdot 
  \vct{p}_{\infty}}{\pin}
  \theta_{\rm cons}
  \bigg)
  +
  .\,.\,. 
\end{align}
For generic masses, this effect starts at 4PM order because the recoil starts at 3PM in the $\vct{p}_\infty$ direction and at 4PM in the $\vct{b}$ direction (the leading-order 1PM part of $\theta_{\rm cons}$ balances the counting).

In the comparisons with NR in this work, we only consider equal-mass scattering scenarios.
In this case, the recoil $\vct{\Delta P}$ is proportional to $\chi_-$ and therefore suppressed by one further PM order (so that $\theta_1-\theta_2$ starts at 5PM order).
This is because, generally,
the recoil is a symmetric function of the two black holes:
its coefficients of $\vct{b}$ and $\vct{\pin}$ must be antisymmetric,
and therefore include either $\delta$ or $\chi_-$.
For the kinematics relevant to the NR comparisons of this work, namely $\nu=1/4$, $\Gamma=1.023$ and $\ell = 4.58$, Eq.~\eqref{eq:angleDifferenceAnalytic} evaluates to:
\begin{align}\label{eq:angleDifference}
  \theta_1^{\rm 5PM}-\theta_2^{\rm 5PM}
  \approx
  -8.4^{\circ}
  \times 10^{-4}
  \chi_-
  \ .
\end{align}
Hence, this effect is extremely small compared, e.g., to the 4PM scattering angle,
which for these kinematics (and to leading order in spins)
takes the value $\theta^{\rm 4PM}\approx144.5^{\circ}+\mO(\chi_\pm)$.
We mostly therefore ignore this effect in our subsequent comparisons of the angle with NR data.\footnote{One remaining puzzle is that the 5PM perturbative result
Eq.~\eqref{eq:angleDifference} suggests $(\theta_1-\theta_2)/\chi_-<0$.
Meanwhile, the NR data of Ref.~\cite{Rettegno:2023ghr} for unequal spins
uniformly has $(\theta_1-\theta_2)/\chi_->0$.
This NR data, however, exists mostly outside the PM-expansion's regime of validity.
It will be important to understand this behavior in the future when more NR data 
will be available.}
We note also that the 5PM effect in Eq.~\eqref{eq:angleDifference} is an order of magnitude smaller than the 4PM effect due to unequal masses,
which at its maximum for $\nu\approx0.17$ (and, as before, $\Gamma=1.023$ and $\ell = 4.58$) takes the value $\theta^{\rm 4PM}_1-\theta^{\rm 4PM}_2\approx \pm 5.1^{\circ}\times 10^{-3}$
with a minus (plus) sign for $m_1>m_2$ ($m_2<m_1$).

With the presence of recoil, one may still define a relative scattering angle which is symmetric in the two BHs.
One possibility involves the covariant impulse $\Delta p_{\rm rel}^\mu$ of the relative momentum:
\begin{align}
  p^\mu_{\rm rel} 
  =
   (0,\vct{p}_\infty) =
  \frac{
    E_2 p_1^\mu-E_1 p_2^\mu
  }{M}
  \ .
  \label{eq:relMom}
\end{align}
This impulse is defined by considering the change of each variable of the right-most side of Eq.~\eqref{eq:relMom} including the energies $E_i$.
The relative scattering angle $\theta_{\rm rel}$ is then defined by: 
\begin{align}\label{eq:relativeAngleDefinition}
  \cos\theta_{\rm rel}
  =
  \frac{
    (\vct{p}_\infty
  +
  \Del\vct{p_{\rm rel}})
  \cdot
  \vct{p}_\infty
  }{
    |\vct{p}_\infty
    +
    \Del\vct{p_{\rm rel}}|
    \pin
  }
  \ .
\end{align}
This angle was first computed at 4PM order in Ref.~\cite{Dlapa:2022lmu} (from where we have adopted the relative subscript),
and the 5PM spin-orbit part in Ref.~\cite{Jakobsen:2023hig}.
This relative scattering angle is defined for dissipative motion including recoil.

In the following comparisons against NR (Sec.~\ref{sec:plots}), we will work with equal masses and spins in which case the recoil vanishes.
Generally, for a vanishing recoil the relative scattering angle of Eq.~\eqref{eq:relativeAngleDefinition} coincides with the (also coinciding) individual angles $\theta_i$.
Dissipative effects, however, may still be non-zero.
More generally, for a non-zero recoil, we are not aware of any simple relationship between the relative angle and the individual angles.

\begin{figure*}
  \centering
  \subfloat{
      \includegraphics[width=.5\textwidth]{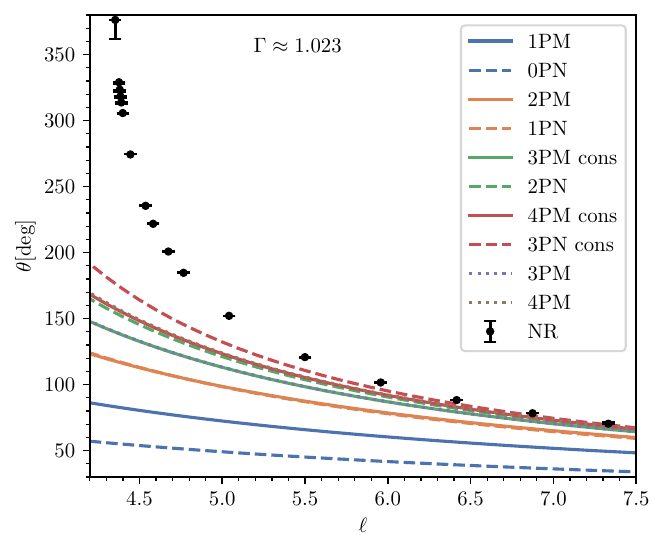}
  }
  \subfloat{
      \includegraphics[width=.5\textwidth]{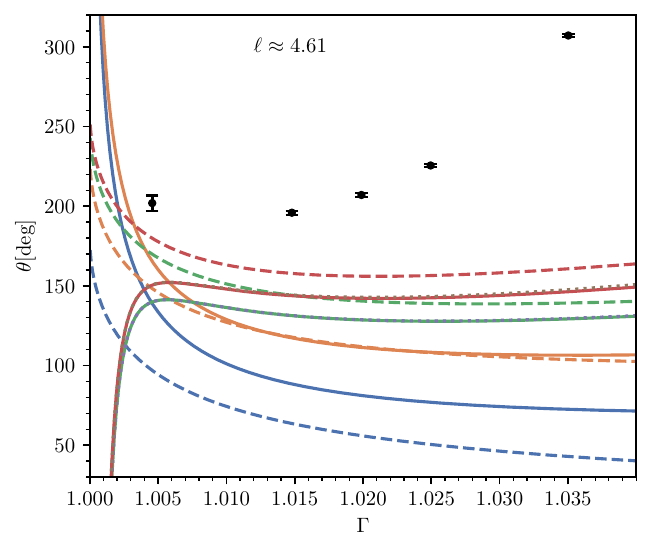}
  }    
  \caption{The perturbative scattering angle $\theta$ plotted at different PM and PN orders, conservative ``cons'' or full dissipative,
      compared with NR-data points~\cite{Hopper:2022rwo,Rettegno:2023ghr}.
      Both plots have equal masses and zero spins,
      absence of recoil at these perturbative orders thus implying $\theta_{\rm rel}=\theta_1=\theta_2$.
      To the left, we fix energy while varying angular momentum;
      to the right vice versa. As pointed out in Ref.~\cite{Damour:2022ybd} for the PM case, the perturbative angles do not perform well when the NR scattering angles become too large (close to the critical angular momentum).
      These results change drastically, and approach the NR data, when evaluating the scattering angle~\eqref{eq:angleFromPR} without first expanding it perturbatively (in PM or PN), thus capturing the pole at the critical angular momentum.}
  \label{fig:vanilla1}
\end{figure*}

We define a split of the relative angle into conservative and dissipative effects by writing:
\begin{align}\label{eq:fullAngle}
  \theta_{\rm rel}
  =
  \theta_{\rm cons}
  +
  \theta_{\rm diss}
  \ .
\end{align}
Since the conservative and relative angles have been defined above in Eqs.~\eqref{eq:conservativeAngle} and~\eqref{eq:relativeAngleDefinition}, the present equation defines the remaining part $\theta_{\rm diss}$ (naturally, when dissipation is negligible, this equation is an identity because then $\theta_{\rm rel}$ and $\theta_{\rm cons}$ coincides).
Similarly to the conservative angle we may determine an effective potential $w_{\rm rel}$ that,
when using Eqs.~\eqref{eq:impetusTrue} and~\eqref{eq:angleFromPR}, reproduces $\theta_{\rm rel}$.
While the physical intuition for this potential is less clear, one can certainly always imagine a conservative system that reproduces a given scattering angle.
This idea was pursued  in Refs.~\cite{Damour:2022ybd,Rettegno:2023ghr}, where it was seen to improve the $w_{\rm PM}$ model drastically (this idea was also suggested in Ref.~\cite{Damgaard:2021rnk}).

  In the PM expansion of the scattering angles, dissipative effects first appear at the 3PM order (without, yet, any recoil effects).
  These are \textit{odd-in-velocity} dissipative effects,
  characterized from the PM loop integration perspective by a single radiative (on-shell) graviton~\cite{Dlapa:2023hsl,Jakobsen:2023hig}.
  Until 4PM order, it was shown that these odd dissipative parts of the scattering angle can be
  reconstructed from the conservative angle through linear response~\cite{Bini:2012ji,Jakobsen:2023hig}.
  At the 4PM order, \textit{even-in-velocity} dissipative effects appear, characterized by two radiative gravitons.
  For similar kinematics as considered above, namely $\nu=1/4$, $\Gamma=1.023$ and $\chi_\pm=0$, one finds that in the PM expansion,
  odd dissipative effects are much larger than the even effects:
\begin{align}\label{eq:OddEvenDiss}
  \frac{
    \theta^{\rm 4PM}_{\rm odd\, diss}
  }{
  \theta^{\rm 4PM}_{\rm even\, diss}
  }
  \approx
  32.2\,
  \ell  
  +
  578.3
  \ .
\end{align}
This is also expected on account of the even effects being suppressed by one PM order in comparison to the odd effects.
Because of the equal masses, there is no ambiguity of the angle in Eq.~\eqref{eq:OddEvenDiss} (i.e., $\theta_1=\theta_2=\theta_{\rm rel}$).

  Let us then discuss the appearance of dissipative effects in the combined PM and PN expansion of the relative angle.
  In Table~\ref{tab:PN}, terms in this expansion where dissipative effects appear are indicated by green shaded cells.
  All dissipative effects shown in this table (i.e. to 5PM and 4PN order) are odd,
  and so appear at a shifted 0.5PN order compared with the corresponding conservative terms.
  Thus, for even orders in spin, odd dissipative effects appear at half-integer PN orders (for odd orders in spin, they appear at integer PN orders).
  Odd dissipative effects appear first at 2.5PN order while even dissipative effects first appear at 5PN order (and thus are not present in Tabel~\ref{tab:PN}).
  Generally, in the PM and PN expansions, dissipative effects first show up in the non-spinning part;
  spin-dependence pushes their appearance to higher perturbative orders.
  From the loop counting perspective, however, dissipative effects appear uniformly at the 2-loop order (see Table~\ref{tab:PMtable}).

  Analytic dissipative and conservative PM results along the rows of Table~\ref{tab:PN} are known except for the spinless 5PM part of the angle (see, however, the recent~\cite{Driesse:2024xad}).
We are, however, not aware of a dissipative result for the PN angle in the sense of Eq.~\eqref{eq:anglePN}, which would resum the 2.5PN spinless column of Table~\ref{tab:PN}.
We note, also, that the dissipative effects of the potential as defined in the present work do not seem to have the same upper triangular pattern as the conservative results.
As the 5PM $S^0$ terms are not yet known, however, we cannot draw any definite conclusions.

Let us finally consider how the perturbative $n$PM and $m$PN scattering angles compare against equal-mass, nonspinning NR data.
In Fig.~\ref{fig:vanilla1} we plot these angles for $1\le n\le4$ and $0\le m\le3$ against NR data from Refs.~\cite{Hopper:2022rwo,Rettegno:2023ghr}\footnote{We note that we could only use the non-spinning NR data in Ref.~\cite{Hopper:2022rwo}, because for the spinning 
data we found an inconsistency between the total angular momentum and the sum of the orbital angular momentum and the binary's total spin.} (the PM curves of the left panel have already appeared in Refs.~\cite{Khalil:2022ylj,Damour:2022ybd}).
For $n\ge3$ and $m\ge3$ we distinguish between the conservative angles (labelled cons) and the dissipative ones (where, again, equal mass means that $\theta_1=\theta_2=\theta_{\rm rel}$).
Clearly, in the left panel, both the higher-order perturbative PM and PN angles describe the NR data
well for sufficiently large angular momentum $\ell$. Generally, however, as noticed in Ref.~\cite{Damour:2022ybd}, close to the critical angular momentum beyond which the plunge occurs, the agreement between the perturbative angles and NR is poor. This is due to the fact that by expanding Eq.~(\ref{eq:angleFromPR}) (with $p_r$ from Eq.~(\ref{eq:impetusTrue})) perturbatively (PN or PM), one cannot capture the pole in the scattering angle corresponding to the critical angular momentum.

Furthermore, in the right panel of Fig.~\ref{fig:vanilla1}, where the energy is varied, one notes a clear difference between the PN and PM angles.
In the low velocity $v\to0$ limit ($\Gam\to1$),
the PN angles have a finite limit while the PM angles diverge.
One can easily read off the $v\to0$ asymptotic behavior with fixed $\ell$ of the $n$PM angles from the (spinless) 0PN column of Table~\ref{tab:PN}:
\begin{align}
  \theta^{2n{\rm PM}}
  \sim
  \theta^{(2n+1){\rm PM}}
  \sim 
  v^{-2n-1}
  \ .
\end{align}
However, after resumming the entire column one finds the result in Eq.~\eqref{eq:anglePN}, namely $  2
\text{arctan}(
1/\ell v
)$, with the finite limit $\pi$ ($180^\circ$).

For the phase space points considered in Fig.~\ref{fig:vanilla1}, dissipative effects of the PM angle are very small.
However, when computing the results by evaluating the scattering angle from Eq.~(\ref{eq:angleFromPR}) without first expanding it perturbatively (in PM or PN), thus capturing the pole at the critical angular momentum, this is no longer the case, as found in the PM case in Ref.~\cite{Damour:2022ybd}.
We note that Eq.~\eqref{eq:OddEvenDiss} which compares odd and even dissipation is valid for the phase space points of the left panel.
We may therefore gather that, for this kinematics, the conservative part is much larger than the odd dissipation which is much larger than the even dissipation: $\theta^{\rm 4PM}_{\rm cons}\gg \theta_{\rm odd\,diss}^{\rm 4PM}\gg \theta_{\rm even\,diss}^{\rm 4PM}$.

\section{Scattering in the Probe Limit}\label{sec:probeAngle}

To prepare for our discussion of the SEOB-PM model,
let us also review the simple case of a non-spinning probe of mass $\mu$,
moving under the influence of a Kerr BH.
In Boyer-Lindquist coordinates $(t,r,\theta,\phi)$,
where $p_\mu=(E_{\rm Kerr},-p_r,-p_\theta,-p_\phi)$,
the (inverse) Kerr metric takes the form (see, e.g., ~\Rcite{Vines:2016unv})
\begin{align}
\begin{aligned}
  g^{\mu\nu}_{\rm Kerr}\partial_\mu\partial_\nu&=
  \frac{\Lambda}{\Delta\Sigma}\partial_t^2-\frac{\Delta}{\Sigma}\partial_r^2-\frac1{\Sigma}\partial_\theta^2\\
  &\qquad-\frac{\Sigma-2Mr}{\Sigma\Delta\sin^2\theta}\partial_\phi^2+\frac{4Mra}{\Sigma\Delta}\partial_t\partial_\phi\,,
\end{aligned}
\end{align}
where we have introduced
\begin{subequations}
\begin{align}
  \Sigma&=r^2+a^2\cos^2\theta\,,\\
  \Delta&=r^2-2Mr+a^2\,,\\
  \Lambda&=(r^2+a^2)^2-a^2\Delta\sin^2\theta\,.
\end{align}
\end{subequations}
Specializing to aligned spins,
we restrict ourselves to the orbital plane $\theta=\sfrac{\pi}2$.
We find it helpful to introduce the specific combinations (as done in Ref.~\cite{Khalil:2023kep}):
\begin{subequations}
\begin{align}
  A^{\rm Kerr}&=\frac{\Delta\Sigma}{\Lambda}=\frac{1-2u+\chi^2u^2}{1+\chi^2u^2(2u+1)}\,,\\
  B_{\rm np}^{\rm Kerr}&=\frac{r^2}{\Sigma}\left(\frac{\Delta}{r^2}-1\right)=\chi^2u^2-2u\,,\\
  B_{\rm npa}^{\rm Kerr}&=-\frac{r^2}{\Sigma\Lambda}\left(\Sigma+2Mr\right)=-\frac{1+2u}{r^2+a^2(1+2u)}\,,
\end{align}
\end{subequations}
where $a=m\chi$.
We can now solve the mass-shell constraint $g^{\mu\nu}_{\rm Kerr}p_\mu p_\nu=\mu^2$
for the radial momentum $p_r$:
\begin{align}\label{eq:probeImpetus}
  p_r^2&=
  \frac1{(1+B_{\rm np}^{\rm Kerr})}\bigg[\frac1{A^{\rm Kerr}}\left(E_{\rm Kerr}-\frac{2MLa}{r^3+a^2(r+2M)}\right)^2\nn\\
  &\qquad-\left(\mu^2+\frac{L^2}{r^2}+B_{\rm npa}^{\rm Kerr}\frac{L^2a^2}{r^2}\right)\bigg]\,.
\end{align}
Alternatively, by solving for the energy $E_{\rm Kerr}=H_{\rm Kerr}(p_r,L,r;a)$,
we obtain a one-body Hamiltonian:
\begin{align}\label{eq:HKerr}
  &H_{\rm Kerr}=\frac{2MLa}{r^3+a^2(r+2M)}\\
  &\quad+\sqrt{A^{\rm Kerr}\left(\mu^2+\frac{L^2}{r^2}+(1+B_{\rm np}^{\rm Kerr})p_r^2
  +B_{\rm npa}^{\rm Kerr}\frac{L^2a^2}{r^2}\right)}\,.\nn
\end{align}
This is our starting point for setting up the SEOB-PM model.

At leading order in $G$, using Eqs.~(\ref{eq:angleFromPR}) and (\ref{eq:probeImpetus}),
one can derive the tree-level scattering angle~\cite{Vines:2017hyw}:
\begin{align}
  \theta_{\rm tree}=\frac{2}{\ell\sqrt{\gamma_{\rm p}^2-1}}
  \frac{2\gamma_{\rm p}^2-1-2\gamma_{\rm p}(\gamma_{\rm p}^2-1)\frac{\chi}{\ell}}
  {1-(\gamma_{\rm p}^2-1)\frac{\chi^2}{\ell^2}}\,,
\end{align}
where $\gamma_{\rm p}=E_{\rm Kerr}/\mu$ and $\ell$ are the dimensionless energy
and angular momentum of the probe.
As was explained in~\Rcite{Vines:2017hyw},
this formula is in one-to-one correspondence with the tree-level scattering
angle between two comparable-mass spinning bodies:
\begin{align}\label{eq:treeAngle}
  \theta_{\rm tree}=\frac{2}{\ell_{\rm cov}\sqrt{\gamma^2-1}}
  \frac{2\gamma^2-1-2\gamma(\gamma^2-1)\frac{\chi_+}{\Gamma \ell_{\rm cov}}}
  {1-(\gamma^2-1)\frac{\chi_+^2}{\Gamma^2\ell_{\rm cov}^2}}\,,
\end{align}
which involves the (dimensionless) covariant angular momentum $\ell_{\rm cov}=L_{\rm cov}/(M\mu)$.
This formula accounts for the entire first row of \Tab{tab:PMtable},
and holds to arbitrarily high orders in spin.
In subsequent work~\cite{Vines:2018gqi}, it was also shown that --- with an appropriate EOB mapping ---
the correspondence may be extended to 2PM order, including spin effects.
We note that in the non-spinning case at 1PM order such a result was obtained in Ref.~\cite{Damour:2017zjx}.

\section{SEOB-PM Hamiltonian and $w$ potential}\label{sec:EOB}

We now derive the main result of this work, a spinning EOB Hamiltonian
based on the PM perturbative results (SEOB-PM) that fully accounts for
hyperbolic motion through 4PM order. We employ it in this work to
compute its corresponding (resummed) conservative and dissipative scattering angles,
while Ref.~\cite{Buonanno:2024byg} uses it to derive waveform
models for spinning BHs on generic orbits upon completing it with the
non-local--in-time 4PN contributions for bound orbits.

The EOB formalism maps the spinning two-body dynamics onto the
dynamics of an effective test-body in a deformed Kerr spacetime. Here,
when including PM contributions into the EOB Hamiltonian (i.e., the
SEOB-PM model), we start from the mass-ratio deformation of the probe
limit for a test mass in Kerr spacetime that was used in
Ref.~\cite{Khalil:2023kep} (see Eq.~(26) therein) to build the
most-recent SEOB model based on PN results (i.e., the {\tt SEOBNRv5}
model~\cite{Pompili:2023tna,Ramos-Buades:2023ehm} used by the LVK Collaboration). Then, we include
PM results by generalizing to the spinning case the so-called
post-Schwarzschild* (PS*) deformation of the geodesic motion
introduced in Refs.~\cite{Antonelli:2019ytb,Khalil:2022ylj}, wherein
$g^{\mu\nu}_{\rm eff}p_\mu p_\nu=\mu^2$.\footnote{We find that the
  alternative post-Schwarzschild (PS) deformation, wherein $g^{\mu\nu}_{\rm Schw}p_\mu
  p_\nu=\mu^2+Q$, with deformations incorporated into
  $Q$~\cite{Damour:2017zjx,Antonelli:2019ytb,Khalil:2022ylj}, gives a
  weaker agreement with NR data for the binding energy for
  bound orbits, and also for scattering trajectories. Therefore we do not
  describe this EOB model.} Thus, we obtain
\begin{align}\label{eq:HSEOBPM}
  &H_{\rm eff}=\frac{ML(g_{a_+}a_++g_{a_-}\delta a_-)}{r^3+a_+^2(r+2M)}\\
  &\qquad+\sqrt{A\left(\mu^2+\frac{L^2}{r^2}+(1+B_{\rm np}^{\rm Kerr})p_r^2
  +B_{\rm npa}^{\rm Kerr}\frac{L^2a_+^2}{r^2}\right)}\,,\nn
\end{align}
where we identify the orbital angular momentum $L$ with the one of the effective test-body, and 
choose the (deformed) Kerr spin $a$ to be $a_+$~\cite{Khalil:2023kep}.
The resummed EOB two-body Hamiltonian is then given by
\begin{align}\label{eq:Heob}
  H_{\rm SEOB-PM}=M\sqrt{1+2\nu\left(\frac{H_{\rm eff}}{\mu}-1\right)}\,,
\end{align}
which is the usual EOB energy map.
As the deformations within $H_{\rm eff}$ (i.e., in $A$ and $g_{a_\pm}$, as seen in
Eq.~\eqref{eq:def} below) are themselves $\gamma$-dependent, the
Hamiltonian technically depends on itself.\footnote{\label{def}To
circumvent this problem, in Refs.~\cite{Damour:2017zjx,Antonelli:2019ytb,Khalil:2022ylj}, 
for the non-spinning case, $\gamma$ was expressed in terms of 
$H_{\rm Schw}$ plus suitable PM corrections, depending on the PM order 
at which the Hamiltonian was computed (see for details Appendix B in Ref.~\cite{Khalil:2022ylj}).}
Therefore, following our discussion in Sec.~\ref{sec:consAngle} and imposing $E_{\rm eff}=H_{\rm eff}$,
we rewrite the above equation in terms of $p_r^2$
(thus placing all $E_{\rm eff}$ dependence on the right-hand side):
\begin{subequations}
\begin{align}
  p_r^2
  &=
  \frac1{(1+B_{\rm np}^{\rm Kerr})}\bigg[\frac1{A}\left(E_{\rm eff}-\frac{ML(g_{a_+}a_++g_{a_-}\delta a_-)}{r^3+a_+^2(r+2M)}\right)^2
  \nn\\
  &\qquad-\left(\mu^2+\frac{L^2}{r^2}+B_{\rm npa}^{\rm Kerr}\frac{L^2a_+^2}{r^2}\right)\bigg]\,,\\
  & \equiv
  \pin^2-\frac{L^2}{r^2}
  +
  w_{\rm SEOB-PM}(E_{\rm eff},L,r;a_\pm). \label{eq:impetusSEOBPM}
\end{align}
\end{subequations}
The last equation (impetus formula) defines a specific resummation of the $w$ potential for the SEOB-PM model, which 
we use in the rest of this work when comparing the scattering angle to NR data~\cite{Hopper:2022rwo,Rettegno:2023ghr} and the $w_{\rm PM}$-potential model~\cite{Damour:2022ybd,Rettegno:2023ghr}.

To build the SEOB-PM effective Hamiltonian~\eqref{eq:HSEOBPM},
we incorporate even-in-spin corrections into the $A$ potential,
and odd-in-spin corrections into the gyro-gravitomagnetic coefficients $g_{a_\pm}$:\footnote{Typically
  gyro-gravitomagnetic factors are taken independent of the spins,
  and thus account for only linear-in-spin corrections to the model;
  we choose to include \emph{all} odd-in-spin terms here,
  and thus avoid introducing further deformation functions.}
\begin{align} \label{eq:def}
  A&=\frac{1-2u+\chi_+^2u^2+\Delta A}{1+\chi_+^2u^2(2u+1)}\,,&
  g_{a_\pm}&=\frac{\Delta g_{a_\pm}}{u^2}\,,
\end{align}
In the non-spinning probe limit $\nu\to0$, where also $a_\pm\to a_1$ (i.e.,~the spin on the probe $a_2$ vanishes), we demand that
$A\to A^{\rm Kerr}$ and $g_{a_+}+g_{a_-}\to2$;
thus, SEOB-PM reduces to the probe limit~\eqref{eq:probeImpetus}.
Deformations of the model are PM-expanded:
\begin{align}
  \Delta A&=\sum_{n\geq 2}u^n\Delta A^{(n)}\,,&
  \Delta g_{a_\pm}&=\sum_{n\geq 2}u^n\Delta g_{a_{\pm}}^{(n)}\,.
\end{align}
As the linear-in-$G$ scattering angle contains only the non-spinning probe limit,
it is already encoded by the undeformed impetus formula~\eqref{eq:probeImpetus};
thus, our deformations begin at quadratic order in $G$.
The $A$-potential incorporates even-in-spin corrections:
\begin{align}
  \Delta A^{(n)}=
  \sum_{s=0}^{\lfloor (n-1)/2 \rfloor}\sum_{i=0}^{2s}
  \alpha^{(n)}_{(2s-i,i)}\delta^{\sigma(i)}\chi_+^{2s-i}\chi_-^{i}\,.
\end{align}
The dimensionless coefficients $\alpha^{(n)}_{(i,j)}$
are functions of the boost factor $\gamma$
(in this context the dimensionless effective energy $E_{\rm eff}$)
and the dimensionless mass ratio $\nu$. We incorporate
odd-in-spin corrections to the model into the gyro-gravitomagnetic factors
$g_{a_\pm}$ (which are themselves even in spin):
\begin{subequations}
\begin{align}
  \Delta g_{a_+}^{(n)}&=\!\!
  \sum_{s=0}^{\lfloor (n-2)/2 \rfloor}\sum_{i=0}^s
  \alpha^{(n)}_{(2(s-i)+1,2i)}\chi_+^{2(s-i)}\chi_-^{2i}\,,\\
  \Delta g_{a_-}^{(n)}&=\!\!
  \sum_{s=0}^{\lfloor (n-2)/2 \rfloor}\sum_{i=0}^s
  \alpha^{(n)}_{(2(s-i),2i+1)}\chi_+^{2(s-i)}\chi_-^{2i}\,.
\end{align}
\end{subequations}
The coefficients $\alpha^{(n)}_{(i,j)}$ are in one-to-one correspondence with the gauge-invariant coefficients
of the full two-body scattering angle ---
except for $\theta^{(1)}_{(0,0)}$, which has no counterpart $\alpha^{(1)}_{(0,0)}$.

The essential constraint on the SEOB-PM model is that the PM-expanded resummed scattering angle
must equal the two-body scattering angle determined from perturbative PM calculations ---
see \Tab{tab:PMtable}:
\begin{align}
  \theta_{\rm SEOB-nPM}=\theta
  +
    \cO\!
    \left(\frac{1}{\ell^{n+1}}\right)\,,
\end{align}
where in general $\theta$ should be identified with $\theta_{\rm rel}$ of Sec.~\ref{sec:dissAngle}, which 
has both conservative and dissipative contributions (see Eq.~\eqref{eq:fullAngle}).
In order to determine the coefficients $\alpha^{(n)}_{(i,j)}$
we compute the scattering angle perturbatively using \Eqn{eq:angleFromPR}.
We simply PM-expand $p_r$, and then perform the $r$-integration.
However, a particular challenge when performing these integrals
is that one encounters divergences;
furthermore, $r_{\rm min}$ must itself also be determined perturbatively.
To solve both problems a convenient solution is to instead use~\cite{Damour:1988mr,Damour:2019lcq}
\begin{align}\label{eq:angleUseful}
  \begin{aligned}
    \theta_{\rm SEOB-PM}
    =-\pi-2{\rm Pf}\int_{\bar{r}_{\rm min}}^\infty\!{\rm d}r\frac{\partial}{\partial L}p_r(E_{\rm eff},L,r;a_\pm)\,,
  \end{aligned}
\end{align}
where $\bar{r}_{\rm min}=M\ell/\sqrt{\gamma^2-1}$ is the turning point only at leading-PM order.
The \emph{partie finie} (Pf) operation instructs us to take only the non-divergent term.

Following this procedure up to 2PM order we find that
\begin{align}
  &\theta_{\rm SEOB-PM}=
  \frac{2(2\gamma ^2-1)}{\ell\sqrt{\gamma ^2-1}}+
  \frac{\pi \big(\gamma ^2 (15-2\alpha^{(2)}_{(0,0)})-3\big)}{4\ell^2}\nn\\
  &-\frac{2 \gamma  \sqrt{\gamma ^2-1}\big(\alpha^{(2)}_{(1,0)}\chi_++\alpha^{(2)}_{(0,1)}\delta\chi_-\big)}{\ell^2}+\cO(\ell^{-3})\,.
\end{align}
No deformations appear at 1PM, and the scattering angle already agrees with the known result~\eqref{eq:angle1PMand0PN}.
At 2PM order, comparing with the expansion of the scattering angle given in \Eqn{eq:angle},
we may straightforwardly invert to yield the deformation coefficients as function of the scattering angle:
\begin{subequations}\label{eq:deformations2PM}
\begin{align}
  \alpha^{(2)}_{(0,0)}&=
  \frac{15}{2}-\gamma^{-2}\bigg(\frac{2\theta^{(2)}_{(0,0)}}{\pi }+\frac{3}{2}\bigg)\,,\\
  \alpha^{(2)}_{(1,0)}&=-\frac{\theta^{(2)}_{(1,0)}}{2 \gamma  \sqrt{\gamma ^2-1}}\,,\,\,
  \alpha^{(2)}_{(0,1)}=-\frac{\theta^{(2)}_{(0,1)}}{2 \gamma  \sqrt{\gamma ^2-1}}\,.
\end{align}
\end{subequations}
Plugging in the known results, we find that
\begin{subequations}
\begin{align}
  \alpha^{(2)}_{(0,0)}&=\frac{3(\Gamma-1)(5\gamma^2-1)}{2\gamma^2\Gamma}\,,\\
  \alpha^{(2)}_{(1,0)}&=\frac{2}{\Gamma}-\frac{(\Gamma-1)(2\gamma^2-1)}{2 \gamma  (\gamma ^2-1) \Gamma  \nu }\,,\\
  \alpha^{(2)}_{(0,1)}&=\frac{(\Gamma -1)(2 \gamma ^2-1)}{2 \gamma  (\gamma ^2-1) \Gamma  \nu }\,.
\end{align}
\end{subequations}
A similar matching of the spin-orbit gyro-gravitomagnetic factors
$g_{a_\pm}$ up to physical 3PM order (1-loop) has previously been performed in
\Rcites{Bini:2017xzy,Bini:2018ywr}, and our results agree precisely 
(see Eqs.~(7.3) and (7.7) of \Rcite{Bini:2018ywr}).

In the probe limit $\nu\to0$, $\delta\to1$ and $\Gamma\to1$ implies that
$\alpha^{(2)}_{(0,0)}\to0$ (i.e.,~the non-spinning deformation vanishes).
As for the spinning coefficients, $\alpha^{(2)}_{(1,0)}+\alpha^{(2)}_{(0,1)}\to2$
reflects the SEOB-PM model being built around the motion of a non-spinning probe (or test mass) 
moving in a Kerr background: $a_2=0$, implying $a_+=a_-$. 
For the higher-PM deformations, we generically observe that
\begin{align}
  \sum_{i=0}^s\alpha^{(n)}_{(i,s-i)}\xrightarrow{\nu\to0}
  \begin{cases}
    2,& \text{if }s=1, n=2\,,\\
    0,& \text{otherwise}\,.
  \end{cases}
\end{align}
\begin{figure*}
  \subfloat{
    \includegraphics[width=.5\textwidth]{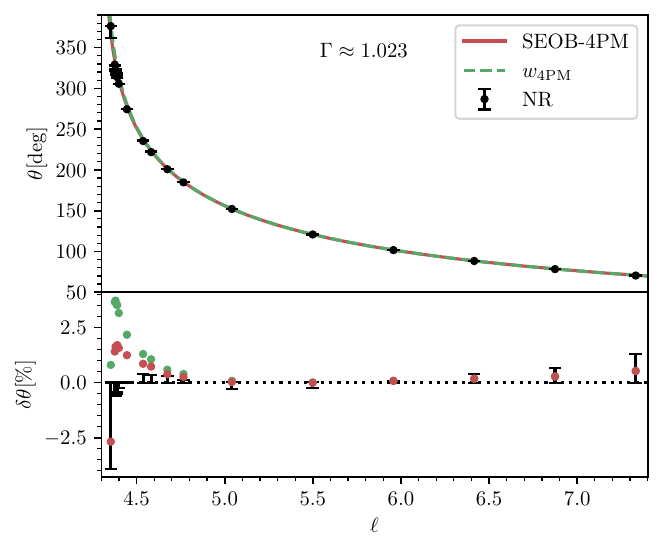}
  }
  \subfloat{
    \includegraphics[width=.5\textwidth]{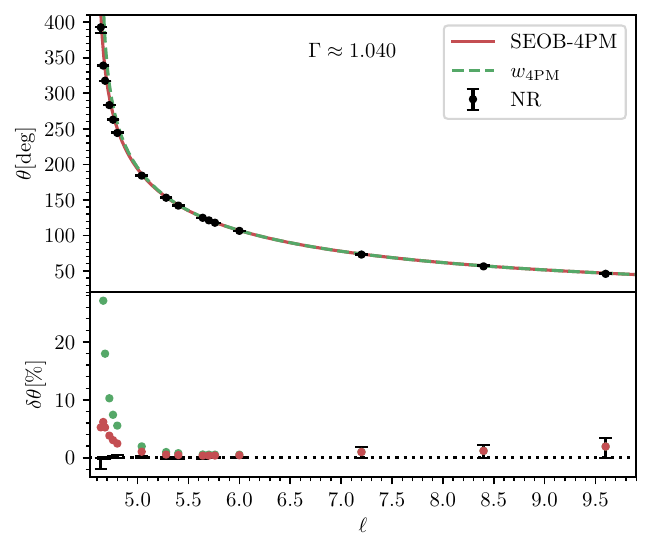}
  }

  \subfloat{
    \includegraphics[width=.5\textwidth]{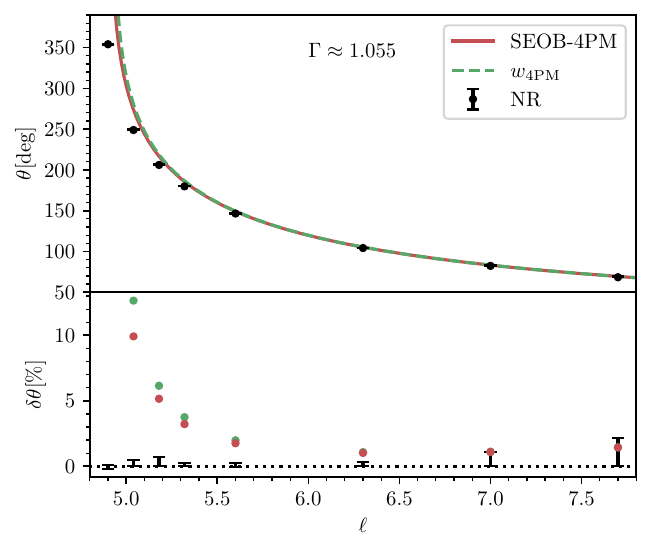}
  }
  \subfloat{
    \includegraphics[width=.5\textwidth]{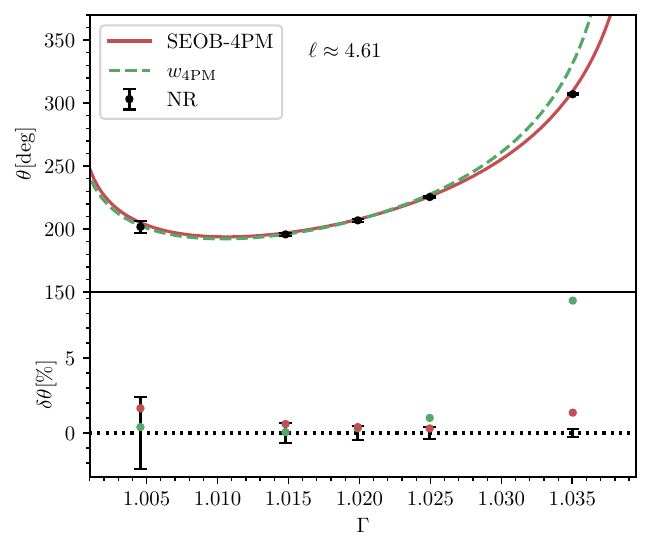}
  }
  \caption{
    Comparison of the dissipative SEOB-4PM and $w_{\rm 4PM}$ models against the four equal-mass spinless NR data sets of Refs.~\cite{Rettegno:2023ghr,Hopper:2022rwo}.
    In each figure, the scattering angle $\theta$ (in degrees) is shown in the top panel,
    and the fractional difference $\delta \theta$~\eqref{eq:relDif} in the lower panel.
    In the bottom-right figure, the energy $\Gam$ is varied while the angular momentum $\ell$ is held fixed;
    vice versa in the other three figures.
    Both models show good agreement with NR, but this worsens in the strong-field regime
    near plunge (small $\ell$).
    The agreement also worsens for large energies with fixed angular momentum,
    particularly for the $w_{\rm PM}$ model. }
    \label{fig:spinlessNR}
\end{figure*}
\begin{figure*}
  \centering
  \subfloat{
  \includegraphics[width=.5\textwidth]{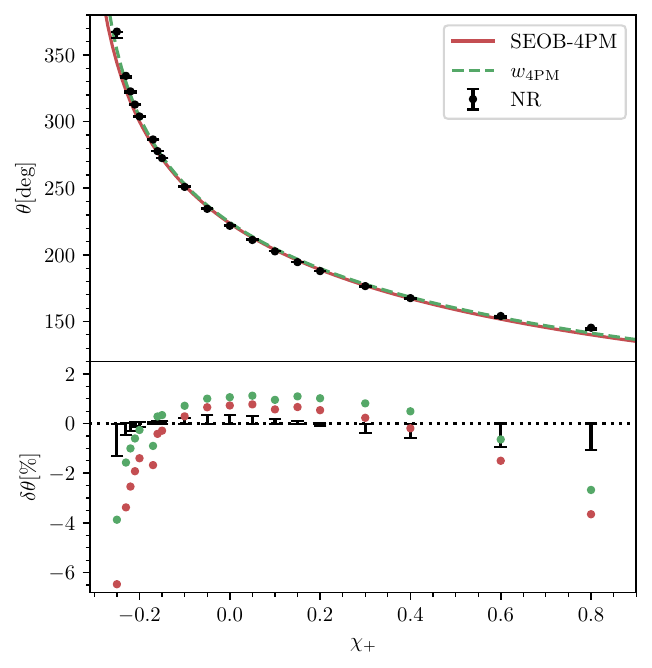}}
  \subfloat{
    \includegraphics[width=.5\textwidth]{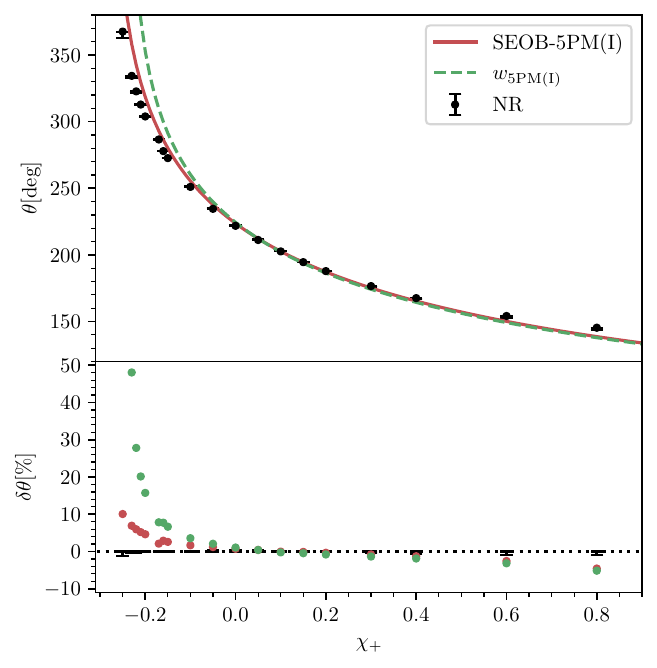}}
  \caption{
    Comparisons with the equal-mass, equal-spins NR simulations of Ref.~\cite{Rettegno:2023ghr}.
    The top and lower panels show the absolute and fractional scattering angles respectively.
    The energy and angular momentum are kept approximately fixed with $\Gamma\approx1.023$ and $\ell\approx4.58$.
    On the left panel we show the SEOB-PM and $w_{\rm PM}$ models at 4PM order, and on the right panel at the incomplete 5PM order, since the non-spinning 5PM 
contributions are currently unknown. }
  \label{fig:spinningNR}
\end{figure*}
Proceeding in this way to higher PM orders, we reconstruct all of the needed
coefficients from the scattering angle.
A complete set of results up to 4PM is provided in \App{sec:deformations},
and up to 5PM (excluding the non-spinning component) in the attached ancillary file. 

Starting at two-loop order (3PM in the non-spinning case),
we may choose to insert either the conservative or full dissipative scattering angle
into the model, as part of this matching procedure.
We will examine both possibilities in the next section.
In either case, the deformations also now include special functions of $\gamma$ that are inherited from the scattering angle:
for example, ${\rm arccosh}\gamma$ when including two-loop results.
At three-loop order (4PM in the non-spinning case), we also encounter logarithms and dilogarithms of rational functions of $\gamma$,
plus the complete elliptic functions ${\rm K}(\sfrac{\gamma-1}{\gamma+1})$ and ${\rm E}(\sfrac{\gamma-1}{\gamma+1})$
of the first and second kind, respectively.

Lastly, we note that in 
the non-spinning limit, the conservative scattering angles of the SEOB-PM model coincide 
with the ones obtained in Ref.~\cite{Khalil:2022ylj}, since our model uses the PS* gauge. On the other hand, 
the dissipative effects differ from the ones of Ref.~\cite{Khalil:2022ylj}, because the latter included 
only the odd contributions following Ref.~\cite{Bini:2012ji}, since the even contributions were absent at the 
time the paper was published. In particular, working 
at linear order in the radiation reaction, the odd radiative contributions to the total scattering angle are 
estimated as half of the difference of the conservative scattering angle evaluated on the 
outgoing and incoming states (see the $H^{\rm EOB, PS*}_{\rm 4PM, hyp}$ model with and without odd dissipation in the right panels of Figs.~7 and 8 in Ref.~\cite{Khalil:2022ylj}). Furthermore, the conservative scattering angle in Ref.~\cite{Khalil:2022ylj} was computed by evolving the EOB Hamilton equations without the radiation-reaction force, and following the substitutions for $\gamma$ explained in the footnote \ref{def}. We will see below the impact of those differences when comparing 
the results of Ref.~\cite{Khalil:2022ylj} with the SEOB-PM model and the NR data.

\section{EOB Scattering Angle and NR Comparisons}
\label{sec:plots}
Let us now consider scattering angle predictions of the SEOB-PM model.
In Sec.~\ref{sec:NRComparisons} we compare these against angles computed from NR simulations of Refs.~\cite{Hopper:2022rwo,Rettegno:2023ghr} including also predictions of the $w_{\rm PM}$ model~\eqref{eq:wEOBModel} of Refs.~\cite{Damour:2022ybd,Rettegno:2023ghr}. Since both models can be expressed in terms of an effective potential $w(r)$ \eqref{eq:impetusTrue}, given explicitly in Eq.~\eqref{eq:impetusSEOBPM} and in Eq.~\eqref{eq:wEOBModel}, their predictions for the scattering angle are then given simply by~\eqref{eq:angleFromPR}:
\begin{align}
  \theta
  =-\pi-2
  \int^\infty_{r_{\rm min}}
  \d r
  \frac{\partial}{\partial L}\sqrt{\pin^2
  -
  \frac{L^2}{r^2}
  +w(E_{\rm eff},L,r;a_\pm)}\,.
\end{align}
As stated above, $r_{\rm min}$ is the largest root of the equation $p_r=0$.
Since the potentials depend on $r$ in quite general manners (in particular the SEOB-PM potential),
the integral cannot easily be evaluated analytically.
Thus, we simply evaluate this integral numerically to a sufficient degree of precision.
The energy, angular momentum, masses and spins are fixed to values corresponding to the particular phase-space point under consideration.

\begin{figure}
  \includegraphics[width=.5\textwidth]{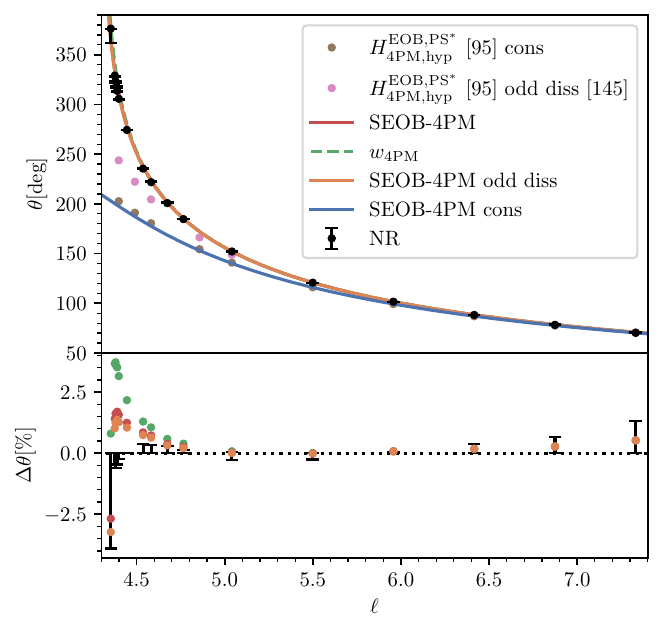}
  \caption{Relative importance of odd and even dissipative effects in the SEOB-4PM model. In addition, we compare the angles' predictions of the SEOB-PM model here and in Ref.~\cite{Khalil:2022ylj}, where odd dissipative effects were incorporated with the radiation reaction of Ref.~\cite{Bini:2012ji}. (We thank Mohammed Khalil for sharing with us the data, including the results of the 
last three data points in the strong field from Ref.~\cite{Damour:2014afa} that were not included in Ref.~\cite{Khalil:2022ylj}.)
\label{fig:khal}
  }
\end{figure}
\begin{figure*}
  \centering
  \subfloat{{
      \includegraphics[width=.5\textwidth]{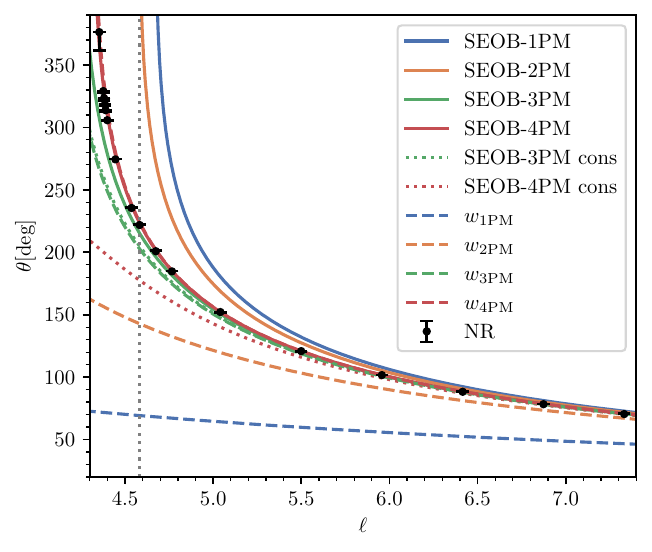}
  }}
  \subfloat{{
      \includegraphics[width=.5\textwidth]{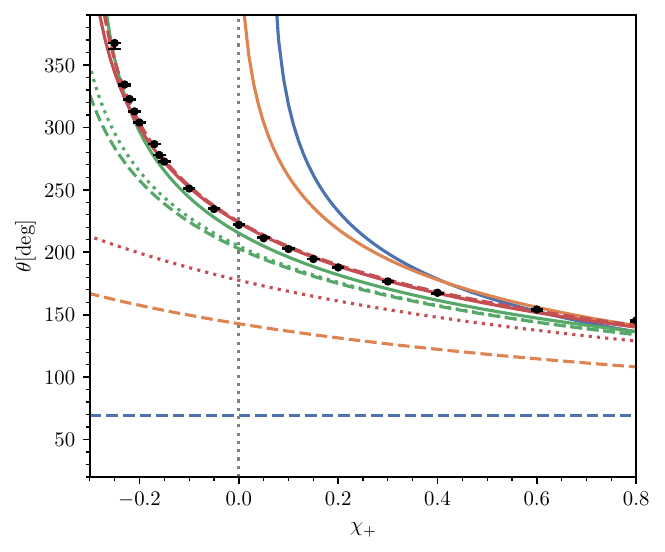}
  }}
  \caption{
    Analysis of the models at different PM orders,
    with or without dissipative effects.
    The energy is fixed $\Gamma\approx1.023$: to the left $\ell$ is varied with fixed $\chi_\pm=0$, while to the right $\chi_+$ is varied with fixed $\ell\approx4.58$ and $\chi_-=0$. The two plots share a single phase-space point marked with the vertical dotted grey lines.
  }
  \label{fig:perturbativePM}
\end{figure*}

Each model comprises a series of models corresponding to the kind of perturbative input given to the deformations.
The most basic series of submodels here are the ones corresponding to a given PM order --- 
where one may choose to include dissipative effects, or not.
Below in Sec.~\ref{sec:perturbativeAndMass}, however, we also explore the PN and spin expansions of these deformations.
Here, we also compare the SEOB-PM and $w_{\rm PM}$ models for unequal masses $\nu\neq1/4$. Finally, in Sec.~\ref{sec:critical}, we compute the critical angular momentum predicted by the SEOB-PM and $w_{\rm PM}$ models and analyze their effective potentials.

\subsection{NR Comparisons}
\label{sec:NRComparisons}

The available NR simulations for the scattering angles of two BHs are still rather limited, and are all restricted to equal masses. We consider here the three non-spinning series of simulations of
Ref.~\cite{Rettegno:2023ghr} with varying angular momentum, and three
fixed energies: $\Gamma\approx1.023$, $\Gamma\approx1.040$ and
$\Gamma\approx1.055$ (similar simulations with the first energy were first carried out in Ref.~\cite{Damour:2014afa}).  We consider also the single spinning series of simulations of Ref.~\cite{Rettegno:2023ghr} with varying (equal)
dimensionless spins and fixed energy $\Gamma\approx1.023$ and angular
momentum $\ell\approx4.58$.  Finally, we consider the single non-spinning
series of simulations of Ref.~\cite{Hopper:2022rwo} with varying
energy and fixed angular momentum $\ell\approx4.61$.

In this section we focus on the models that include dissipation and compare those 
with the NR simulations.  Thus, 4PM is the highest order at which the
perturbative data is known and, also, the dissipative models agree
much better than the conservative models with the NR data.  The 4PM
dissipative models explored here are therefore the most accurate models
available.  We explore the conservative part of the models below
in Sec.~\ref{sec:perturbativeAndMass}.

In Fig.~\ref{fig:spinlessNR}, we compare the non-spinning SEOB-PM and $w_{\rm PM}$ models, at 4PM with dissipative effects, against the non-spinning NR data. Comparisons to the NR data for the $w_{\rm PM}$ model have already appeared in Refs.~\cite{Damour:2022ybd,Rettegno:2023ghr} and our results are in full agreement. In each plot we show the SEOB-4PM and $w_{\rm 4PM}$  predictions across the relevant ranges of angular momentum or energy together with the NR simulation data points.
In most of the phase space shown, the two models and the NR data lie very closely. 
In order to distinguish their behaviour, we also plot in each case the fractional difference $\delta \theta$:
\begin{align}\label{eq:relDif}
  \delta \theta
  =
  \frac{\theta_{\rm model}-\theta_{\rm NR}}{\theta_{\rm NR}}
  \ ,
\end{align}
where $\theta_{\rm model}$ is the angle computed either with the SEOB-4PM or $w_{\rm 4PM}$ models.

Generally the agreement between the models and NR data is rather remarkable: only in the strong field, near plunge does the relative difference rise to more than $10\%$. For the SEOB-4PM model, it never goes beyond $10\%$.
When the models predict a plunge rather than scattering, no value for $\delta \theta$ is plotted.
The performance of both the SEOB-4PM and $w_{\rm 4PM}$ models are good and comparable, though the relative difference of 
the SEOB-4PM model generally is smaller than that of $w_{\rm 4PM}$.  This is evident in particular for the series of data at 
fixed angular momentum and varying energy in the bottom-right panel of Fig.~\ref{fig:spinlessNR}.

In Fig.~\ref{fig:spinningNR}, we turn to the equal-mass, equal-spins
simulations with constant energy and angular momentum, but varied spin.
Here, we consider both the 4PM models (in the left panel) and \emph{incomplete} 5PM models (5PM(I)) in the right panel).
The incomplete 5PM model is defined by including all known information at 5PM (i.e. everything except the 4-loop spinless contribution, see Table \ref{tab:PN}).

Considering first the 4PM models in the left panel of Fig.~\ref{fig:spinningNR},
both models agree with the NR simulations within a relative difference of about $5\%$.  It is, however, interesting that the
agreement of both models worsens when the spin is increased.  For positive spins, 
the scattering angle also decreases and one might have expected the models
to perform better in this more perturbative regime (smaller angle).
However, the positive spin becomes quite large, thus it seems
desirable to improve the models in this phase space where the angles
are not too big and should be describable.
This could be achieved by including higher-spin terms beyond 4PM order
(e.g., the tree-level $S^4$ or one-loop $S^3$) or exploring alternative deformations of the Kerr metric. 

Considering, then, the incomplete 5PM models in the right panel of Fig.~\ref{fig:spinningNR}, we see that the agreement of the $w_{\rm PM}$ model with NR is significantly worse, reaching $50\%$. This, to a much lesser extent, is also true 
for the SEOB model, for which the difference with NR rises to about $10\%$.
It will be interesting to see if the models improve once the
genuine non-spinning 5PM contribution is computed. 
(We note that Ref.~\cite{Rettegno:2023ghr} improved the $w_{\rm PM}$ model by fitting the 5PM $S^0$ and 6PM $S^1$ terms to the NR scattering-angle data.)

\subsection{Dependence on Dissipation, Perturbative Orders and Mass Ratio}
\label{sec:perturbativeAndMass}

\begin{figure}
  \centering
  \includegraphics[width=.5\textwidth]{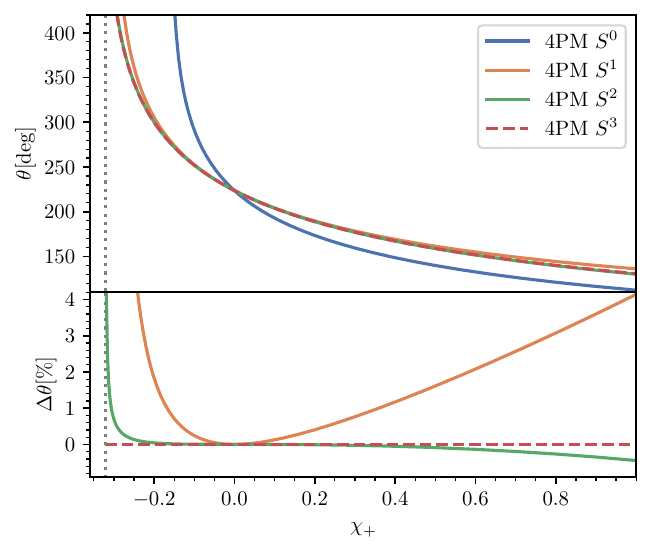}
  \caption{Analysis of importance of perturbative spin orders for the dissipative SEOB-4PM model.
    The kinematics is the same as the equal-mass, equal-spin NR simulations: $\Gamma\approx1.023$ and $\ell\approx 4.58$.
    In the upper panel we plot SEOB-4PM angles omitting first all spin deformations (blue) and adding, then, successively linear (orange), quadratic (green) and cubic (red) spin deformations to the model.
    Starting from linear order in spins, the curves lie very near to each other and in the lower panel we plot their fractional difference to the (full) SEOB-4PM $S^3$ model.
    The vertical dotted line indicates the transition to plunge as predicted by the SEOB-4PM model.
    We note that this figure only compares its submodels, where certain spin deformations are omitted, with itself and not with the NR data.
  }
  \label{fig:perturbativeSpin}
\end{figure}
\begin{figure}
  \centering
    \includegraphics[width=.5\textwidth]{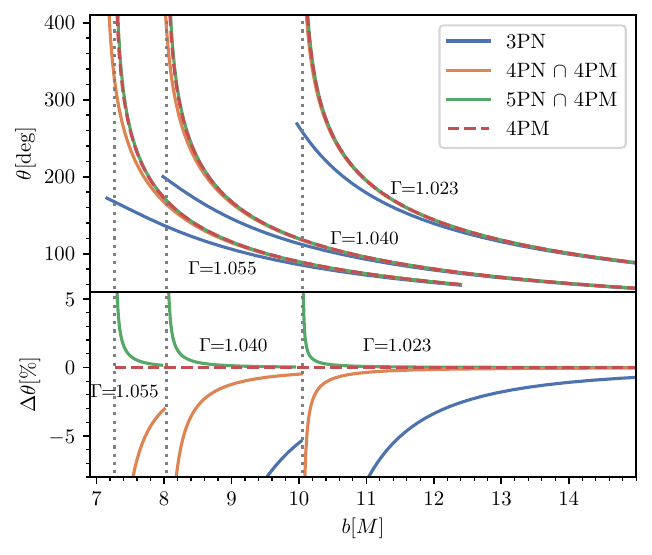}
  \caption{
    Analysis of importance of perturbative PN orders for the (dissipative) SEOB-PM model.
    Focusing on spinless, equal-mass dynamics, we consider three fixed energies $\Gam=1.023$, $1.040$ and $1.055$ (corresponding to the velocities $v\approx0.40$, $0.51$ and $0.58$) and vary the impact parameter (which is used instead of $\ell$ in order that the curves do not lie on top of each other).
    In the lower panel, the fractional difference of each PN approximation with respect to the 4PM model is shown.
    The vertical grey lines indicate the transition to plunge predicted by the 4PM model.
  }
  \label{fig:perturbativePN}
\end{figure}

Let us now analyze the importance of dissipation, and different PM, PN and spin orders.
In Fig.~\ref{fig:khal} we compare the scattering angles of the full SEOB-4PM 
model (i.e., conservative plus odd and even dissipative) with the ones obtained by including 
only the odd dissipative terms. In accordance with Sec.~\ref{sec:dissAngle} (see also Eq.~(\ref{eq:OddEvenDiss})), 
we find that, for the phase-space configurations for which we have NR data, the 
contribution to the scattering angle of the even dissipative terms is negligible, becoming 
more noticeable only in the strong field (see lower panel). We also compare the conservative scattering 
angles of the SEOB-4PM model with the ones of the EOB model of Ref.~\cite{Khalil:2022ylj} 
(see the $H^{\rm EOB, PS*}_{\rm 4PM, hyp}$ model in Figs.~7 and 8 therein), which employed the 
same non-spinning PM Hamiltonian of this work, but computed the angles 
evolving the EOB Hamilton equations without the radiation-reaction 
force (see also the footnote \ref{def}). The two conservative predictions are very close. 
We also show the curve of Ref.~\cite{Khalil:2022ylj} where the odd dissipative 
effects were included, following the method of Ref.~\cite{Bini:2012ji}. 
(The even dissipative terms were not available at that time, and were later
computed in Ref.~\cite{Dlapa:2022lmu}.) We find that the estimation of the odd 
dissipative terms of Ref.~\cite{Bini:2012ji} largely underestimate, at least for this phase-space configurations, 
the true odd dissipative contributions.
 
In Fig.~\ref{fig:perturbativePM} we plot scattering angle predictions of
the SEOB-PM and $w_{\rm PM}$ models at 1, 2, 3 and 4PM orders.  In
addition, at 3PM and 4PM --- where the conservative and dissipative
models differ --- we also plot the conservative SEOB-PM model
predictions.  Clearly, the dissipative effects have a large impact, as
was already observed for the $w_{\rm PM}$
model~\cite{Damour:2022ybd,Rettegno:2023ghr}. 
The dissipative 3PM and 4PM models are, however,
sufficiently close to each other that one might hope for some sort of convergence.
Interestingly, the SEOB-PM model converges from above in both cases
(with an oscillatory behavior between the 3PM and 
4PM order for the SEOB-PM model), in contrast to the $w_{\rm PM}$ which converges from below.
Also, recall that the SEOB-PM model at 1PM encodes simply the probe limit.
We note also that the $w_{\rm 1PM}$ curves may be computed with the 
formula given above in Eq.~\eqref{eq:weob1PM}.

Next, in Fig.~\ref{fig:perturbativeSpin}, we consider the importance of perturbative spin orders.
Focusing on the SEOB-PM model at 4PM order, we omit progressively different orders of the spin corrections from the deformations.
Thus, the models shown labeled by 4PM $S^s$ with $s=0,1,2,3$ denote a model where we include deformations $\alpha^{(n)}_{(i,j)}$ only with $n\le 4$ and $(i+j)\le s$.
Thus, referring to Table~\ref{tab:PMtable}, the model $s=0$ corresponds to the first column, the model $s=1$ corresponds to the first two columns, and so on up to 4PM order.
The top panel shows the scattering angle and the lower panel shows fractional errors with respect to the genuine 4PM model (i.e., 4PM $S^3$). We define the fractional difference in this case by:
\begin{align}
  \label{eq:relativeAngle}
  \Delta \theta
  =
  \frac{\theta_{\rm model}-\theta_{\rm SEOB-4PM}}{\theta_{\rm SEOB-4PM}},
\end{align}
where the subscript ``model'' could be any of the models 4PM $S^s$.
From Fig.~\ref{fig:perturbativeSpin}, it is clear that the spin-orbit $S^1$ corrections
to the model are essential, while the contributions 
from higher spin orders are comparably much smaller. For larger values of the dimensionless spins, however, they do become relevant.

Let us then turn to PN contributions, and ask: is the
all-order-in-$v$ PM information important or can a PN model describe
the NR data equally well?  Again, we focus on the SEOB-PM model and
define a series of sub-models each of which contains only part of the
information of the full 4PM model.  Namely, we terminate the
deformation parameters at a given PN order starting from 3PN and
progressing to 5PN.  Thus, generally, we may define a model
$m\text{PN}\cap 4\text{PM}$ which includes all deformations until
$m$PN order and 4PM order. Referring back to the right panel in 
Table~\ref{tab:PN}, this corresponds to including all terms in the
rectangle extending downwards to 4PM and to the right to $m$PN.  We start
from the model $3\text{PN}\cap4\text{PM}=3\text{PN}$ and include the
next two sub-leading orders in the comparison.  Again, we compute a
relative angle by comparing to the full 4PM model predictions just as
in Eq.~\eqref{eq:relativeAngle}. Fig.~\ref{fig:perturbativePN} shows
the comparison of the $m\text{PN}\cap 4\text{PM}$ against the 4PM
model for $m=3,4,5$.  Only at the sub-sub-leading order to 3PN does
the velocity-expanded models keep in good agreement with the 4PM model
all the way in the strong field, near to plunge.  In the three data sets plotted in
Fig.~\ref{fig:perturbativePN}, one also notes that increasing the
energy worsens the PN approximations.

\begin{figure*}
  \centering
  \subfloat{{
      \includegraphics[width=.5\textwidth]{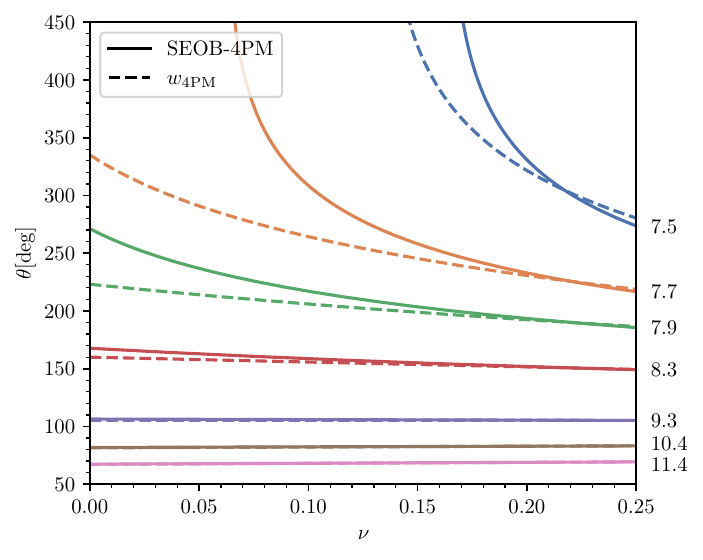}
  }}
  \subfloat{{
      \includegraphics[width=.5\textwidth]{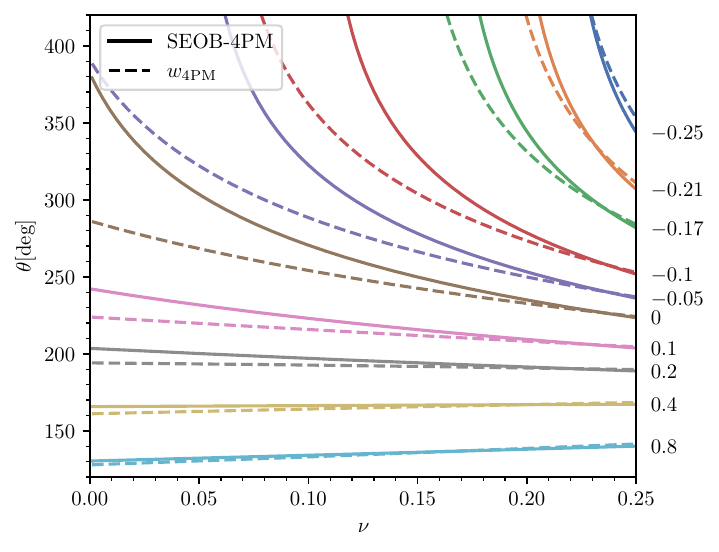}
  }}
  \caption{
    Predictions of the SEOB-4PM and $w_{\rm 4PM}$ models across the full range of symmetric mass ratios $0\le \nu\le1/4$.
    In each curve, we keep the dimensionless effective energy, impact parameter and spins fixed.
    In the left panel, $\gamma\approx1.228$, $\chi_\pm=0$ and the numbers along the right vertical axis indicate values of $b/M$ for the different curves.
    In the right panel, $\gamma\approx1.092$, $b/M\approx10.7$, $\chi_-=0$, and the numbers along the right vertical axis indicate values of $\chi_+$. Note that for unequal masses there is recoil, and thus a difference between $\theta_1$, $\theta_2$ and $\theta_{\rm rel}$.
    The curves shown here are modelled around $\theta_{\rm rel}$ (as discussed in Sec.~\ref{sec:dissAngle}).
    }
  \label{fig:massRatio}
\end{figure*}
\begin{figure}
  \centering
    \includegraphics[width=.5\textwidth]{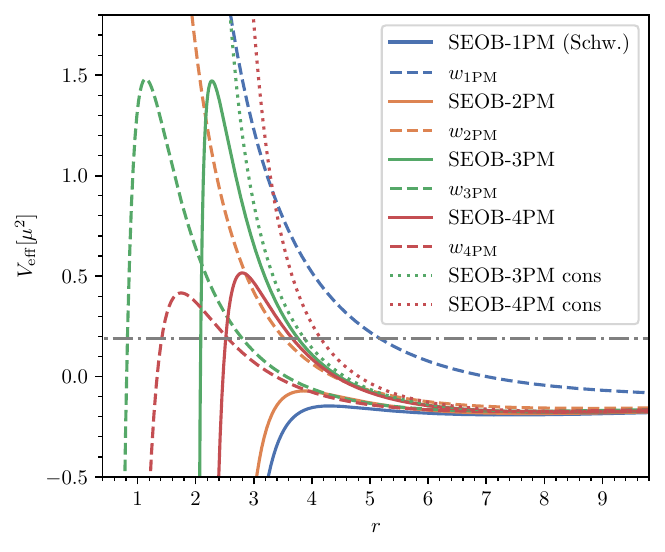}
  \caption{
  For each model we plot the effective potential $V_{\rm eff}(r)$ defined from the radial action as $p_r^2=\pin^2-V_{\rm eff}(r)$.
    The phase space point is equal-mass, non-spinning with $\Gam=1.02264$ and $\ell=4.4$.
    The grey dash-dotted straight line indicates the corresponding value of $\pin^2$:
    a model predicts plunge if the potential never rises beyond that line.
    The effective potentials depend on the kinematics and the condition for the critical angular momentum is that the peak of the potential (if it exists) touches the line of $\pin^2$.
  }
  \label{fig:potentials}
\end{figure}

Finally, we consider the mass dependence of the two models SEOB-PM and $w_{\rm PM}$.
Naturally, the SEOB-PM model is designed to describe exactly the probe limit $\nu\to0$ and $a_+=a_-$, which is a feature not included in the $w_{\rm PM}$ model. In Fig.~\ref{fig:massRatio} we plot angle predictions of the two models across the whole range of $0\le\nu\le1/4$. We do so for the phase-space points of two of the NR data sets.
First, in the left panel, we do so for each value of $\ell$ of the third series of NR data of Ref.~\cite{Rettegno:2023ghr} with energy $\Gamma\approx 1.055$.
Second, in the right panel, we do so for a selection of the equal-spins simulations of Ref.~\cite{Rettegno:2023ghr}.
Note, however, that as a function of $\nu$, we do not keep $\Gamma$ and $\ell$ fixed but instead fix the dimensionless effective energy $\gamma$ and impact parameter $b/M$.
Thus, the $\nu\to0$ limit with fixed total energy implies an arbitrarily large relative velocity $(\gam\to\infty)$ and we rather keep the relative velocity constant.

As we already saw above, the two models give relatively similar results in the $\nu=1/4$ regime.
As seen in Fig.~\ref{fig:massRatio} this, however, is generally not the case for smaller mass ratios.
Thus, for increasing scattering angles, the two models begin to differ more and more for mass ratios $\nu\neq1/4$. 
By design, the leftmost $\nu\to0$ prediction of the SEOB-PM model in the left panel of Fig.~\ref{fig:massRatio} is exact.
This is not the case for the right panel, as that would require a vanishing spin on the probe. 
As an example, for $b/M\approx7.7$ in the left panel, $w_{\rm PM}$ predicts a finite angle in this limit while probe motion would predict a plunge. 

On the other hand, for smaller scattering angles the curves are more or less insensitive to the changing mass ratio. This is in good agreement with the fact that the 1PM and 2PM perturbative scattering angles essentially are independent of $\nu$ (except for too large energies). It will be very important to produce NR simulations 
of scattering BHs for a variety of mass ratios and spins, so that those models can be validated much more 
broadly.
We also remark that for the unequal masses considered in Fig.~\ref{fig:massRatio}, recoil effects will be non-zero and the scattering angles $\theta_1$, $\theta_2$ and $\theta_{\rm rel}$ are different.
While we have no NR data to compare with, one may think of the predictions shown there as referring to $\theta_{\rm rel}$ (which can also be extracted from NR simulations if the impulses $\Delta p_i$ are known).

\subsection{Importance of the Critical Angular Momentum}\label{sec:critical}

An essential feature of the non-perturbative dynamics is the possibility of a plunge for
small values of the angular momentum $\ell$.
Thus, if one fixes all variables but $\ell$, there is a critical value denoted by $\ell_0$ for which the orbital motion goes from scattering to plunge. This scenario is ideally explored in the three panels of Fig.~\ref{fig:spinlessNR} with fixed energy (using the NR data of Ref.~\cite{Rettegno:2023ghr}).
Here, the NR data explores this limit where the curve meets a vertical asymptote (i.e. $\ell_0$).
In fact, every series of NR data was terminated only after a plunge was ascertained for a given value of $\ell$ (see, e.g., the tables in Appendix~\ref{sec:tables}).
In each of the three cases, with the energies $\Gam_1=1.02264$, $\Gam_2=1.04033$ and $\Gam_3=1.05548$, one may therefore bound the critical angular momentum:
\begin{subequations}
  \label{eq:bounds}
\begin{align}
  4.3076
  &\le
  \ell_0(\Gam_1) 
  \le
  4.3536
  \ ,
  \\
  4.602
  &\le
  \ell_0(\Gam_2) 
  \le
  4.638
  \ ,
  \\
  4.2
  &\le
  \ell_0(\Gam_3) 
  \le
  4.9
  \ .
\end{align}  
\end{subequations}
The bound for the third energy, however, is very wide.

The accuracy of an analytical model in the strong field greatly depends
on its ability to predict the critical angular momentum.
Its appearance may be gathered from the shapes of the effective potentials $V_{\rm eff}=\pin^2-p_r^2$ of the models plotted in Fig.~\ref{fig:potentials}.
These shapes are characteristic of the effective potential of BH metrics (i.e., the SEOB-1PM curve in Fig.~\ref{fig:potentials}).
The plunge (or inspiral) happens when there is no longer a barrier generated by the potential.
In other words, when the grey dash-dotted line is never crossed by the potentials.
The critical point of transition from scattering to plunge is then determined
by the potential touching the grey dash-dotted line just once. 
In other words, $p_r(r_{\rm min})=0$ and $\partial p_r(r_{\rm min})/\partial r=0$.
For the three energies $\Gam_i$ we determine the critical angular momenta predicted by the dissipative SEOB-4PM and $w_{\rm 4PM}$ models to be:
\begin{subequations}
\begin{align}
  \ell_0^{\rm SEOB-4PM}(
    \Gamma_1
  )
  &=
  4.3046,
  &
  \ell_0^{\rm wPM}(
    \Gamma_1
  )
  =
  4.3148,
  &
  \\
  \ell_0^{\rm SEOB-4PM}(
    \Gamma_2
  )
  &=
  4.6089,
  &
  \ell_0^{\rm wPM}(
    \Gamma_2
  )
  =
  4.6439,
  &
  \\
  \ell_0^{\rm SEOB-4PM}(
    \Gamma_3
  )
  &=
  4.9072,
  &
  \ell_0^{\rm wPM}(
    \Gamma_3
  )
  =
  4.9272,
  &
\end{align}
\end{subequations}
Generally, these values lie within or near the bounds given in Eqs.~\eqref{eq:bounds}.

\section{Conclusions}
\label{sec:concl}

Building on earlier work~\cite{Bini:2017xzy,Bini:2018ywr,Antonelli:2020ybz,Antonelli:2019ytb,Khalil:2022ylj,Khalil:2023kep},
we have derived the spinning EOB Hamiltonian SEOB-PM that resums PM perturbative calculations fully through 4PM order,
and at 5PM where results are known.
Our PM counting is a physical one, with both loop and spin orders contributing;
thus, both three-loop spin-orbit~\cite{Jakobsen:2023ndj,Jakobsen:2023hig}
and two-loop spin-squared~\cite{Jakobsen:2022fcj,Jakobsen:2022zsx,FebresCordero:2022jts}
scattering results contribute to the model at 5PM order. 
The SEOB-PM Hamiltonian includes nonlocal-in-time (tail) contributions for unbound orbits,
and thus fully describes hyperbolic trajectories.
We have employed the SEOB-PM model to compute resummed conservative scattering angles for non-spinning and spinning BHs, and, after accounting for dissipative contributions, we have compared 
the total scattering angles to the NR data of Refs.~\cite{Hopper:2022rwo,Rettegno:2023ghr}. We have also compared SEOB-PM results 
with the $w_{\rm PM}$-potential--model  predictions in those papers (therein referred to as $w_{\rm EOB}$),
which can be viewed as a PM-expanded version of our $w_{\rm SEOB-PM}$-potential model.

We find that the performance of both the SEOB-PM and $w_{\rm PM}$ 
models are very good and comparable, though the fractional difference
of the SEOB-PM model to NR is generally slightly smaller than that
of $w_{\rm PM}$. This is evident in particular for the set of
scattering angles at fixed angular momentum and varying energy, toward larger energy, 
and also when comparing to NR data the models at the (incomplete) 5PM order.
However, the NR data here is so limited that we cannot draw from
these few examples any definitive conclusions. In fact, in the region
of parameter space where we expect the two models to differ the most, notably
toward the probe limit (i.e. symmetric mass ratio different from 1/4), we do
not have any NR data. Nevertheless, we stress that whereas the SEOB-PM
model (by construction) reduces to the $w$-potential and Hamiltonian
of a test-mass in the Schwarzschild or Kerr spacetime, this is not
the case for the $w_{\rm PM}$ model.

Similarly to what was found for the $w_{\rm PM}$ model in Ref.~\cite{Rettegno:2023ghr}, when comparing 
to spinning NR data for equal-mass BHs, SEOB-PM performs worse when spins are aligned with the angular 
momentum and the spin magnitude increases. Although in this case, the scattering angles are small, thus 
in the weak-field region, the models are not yet sufficiently accurate. This motivates 
the need to complete $G^5$ by computing the non-spinning 5PM scattering dynamics.
We also found that including radiative (dissipative) effects is also necessary for achieving a good agreement with NR;
though, the odd dissipation plays a significantly more important role than the even one. 

The resummed conservative scattering-angle results that we derived here were obtained from an EOB Hamiltonian. Thus, our SEOB-PM model has the advantage that 
it can be tested also for bound orbits and it can be used to construct waveform models. Indeed, Ref.~\cite{Buonanno:2024byg} is already employing the SEOB-PM Hamiltonian at 4PM, augmented with known 
local-in-time contributions at 4PN order for bound orbits~\cite{Damour:2016abl,Marchand:2017pir,Foffa:2019rdf,Foffa:2019yfl,Blumlein:2020pog}, to produce waveform models for spinning BHs on quasi-circular orbits.

\section*{Acknowledgments}
We thank Zvi Bern, Jitze Hoogeveen, Mohammed Khalil, Raj Patil, Jan Plefka, Lorenzo Pompili, and Jan Steinhoff for valuable discussions,
comments and/or work on related projects.
We also thank Mohammed Khalil for a careful reading of this manuscript and insightful comments.
G.J.'s and G.M's work is funded by the Deutsche Forschungsgemeinschaft
(DFG, German Research Foundation)
Projektnummer 417533893/GRK2575 ``Rethinking Quantum Field Theory''.

\appendix

\newpage
\begin{widetext}
\section{SEOB-PM  Deformation Coefficients}\label{sec:deformations}

\noindent
In this appendix we present the deformations required
to fully specify the SEOB-PM model up to physical 4PM order,
as a function of the scattering angle~\eqref{eq:angle}.
No deformations are required at 1PM order;
results up to 2PM are provided in the main text~\eqref{eq:deformations2PM}.
At 3PM, we require results up to quadratic in spin:
\small
\begin{subequations}
  \begin{align}
    \alpha^{(3)}_{(0,0)}&=
    \frac{12 \left(6 \gamma ^4-11 \gamma ^2+5\right) \theta^{(2)}_{(0,0)}+\pi  \left(-142 \gamma ^6+309 \gamma ^4-3 \gamma ^2 (\theta^{(3)}_{(0,0)}v_\infty+68)+3 \theta^{(3)}_{(0,0)}v_\infty+35\right)}{6 \pi  \gamma ^2 v_\infty^4}\,,\\
    \alpha^{(3)}_{(1,0)}&=\frac{\pi  \left(5 \gamma ^2-3\right) \theta^{(2)}_{(1,0)}-2 \theta^{(3)}_{(1,0)} v_\infty}{2 \pi  \gamma  v_\infty^3}\,,\qquad
    \alpha^{(3)}_{(0,1)}=\frac{\pi  \left(5 \gamma ^2-3\right) \theta^{(2)}_{(0,1)}-2 \theta^{(3)}_{(0,1)} v_\infty}{2 \pi  \gamma  v_\infty^3}\,,\\
    \alpha^{(3)}_{(2,0)}&=\frac{2\gamma^2-1}{\gamma ^2}-\frac{\theta^{(3)}_{(2,0)}}{2 \gamma ^2 v_\infty}\,,\qquad
    \alpha^{(3)}_{(1,1)}=-\frac{\theta^{(3)}_{(1,1)}}{2 \gamma ^2 v_\infty}\,,\qquad
    \alpha^{(3)}_{(0,2)}=-\frac{\theta^{(3)}_{(0,2)}}{2 \gamma ^2 v_\infty}\,,
  \end{align}
\end{subequations}
\normalsize
where $v_\infty=\sqrt{\gamma^2-1}$.
At 4PM our results go up to cubic order in spins:
\small
\begin{subequations}
\begin{align}
  \alpha^{(4)}_{(0,0)}&=\frac1{48\pi^2\gamma ^4v_\infty^6}
  \bigg(96 v_\infty^4 \left(3 \gamma ^2-2\right) (\theta^{(2)}_{(0,0)})^2
  -32 \pi  v_\infty^2 \left(2 v_\infty^2 \gamma ^2 \theta^{(4)}_{(0,0)}+3 \left(33 \gamma ^6-58 \gamma ^4+30 \gamma ^2-3\right) \theta^{(2)}_{(0,0)}\right)\\
  &
  +\pi ^2 \left(4127 \gamma ^{10}-12976 \gamma ^8+42 \gamma ^6 (4 \theta^{(3)}_{(0,0)} v_\infty+365)-4 \gamma ^4 (72 \theta^{(3)}_{(0,0)} v_\infty+2011)+\gamma ^2 (120 \theta^{(3)}_{(0,0)} v_\infty+1703)-108\right)\bigg)\nn\,,\\
  \alpha^{(4)}_{(1,0)}&=
  \frac{4 \left(5 \gamma ^4-7 \gamma ^2+2\right) \theta^{(2)}_{(0,0)} \theta^{(2)}_{(1,0)}-2 \pi  v_\infty^2 \gamma ^2 \theta^{(4)}_{(1,0)}+\pi  \left(-99 \gamma ^6+132 \gamma ^4-57 \gamma ^2+6\right) \theta^{(2)}_{(1,0)}+24 \left(2 \gamma
   ^2-1\right) \gamma ^2 \theta^{(3)}_{(1,0)} v_\infty}{8 \pi  \gamma ^3 v_\infty^5}\,,\\
  \alpha^{(4)}_{(0,1)}&=
  \frac{4 \left(5 \gamma ^4-7 \gamma ^2+2\right) \theta^{(2)}_{(0,0)} \theta^{(2)}_{(0,1)}-2 \pi  v_\infty^2 \gamma ^2 \theta^{(4)}_{(0,1)}+\pi  \left(-99 \gamma ^6+132 \gamma ^4-57 \gamma ^2+6\right) \theta^{(2)}_{(0,1)}+24 \left(2 \gamma
   ^2-1\right) \gamma ^2 \theta^{(3)}_{(0,1)} v_\infty}{8 \pi  \gamma ^3 v_\infty^5}\,,\\
  \alpha^{(4)}_{(2,0)}&=
  -\frac{3 \pi  \left(v_\infty \left(84 \gamma ^6-152 \gamma ^4+\gamma ^2 \left(68-15 (\theta^{(2)}_{(1,0)})^2\right)+3 (\theta^{(2)}_{(1,0)})^2\right)+8 \left(5-7 \gamma ^2\right) \gamma
   ^2 \theta^{(3)}_{(2,0)}\right)+64 \gamma ^2 \theta^{(4)}_{(2,0)}v_\infty}{48 \pi  \gamma ^4 v_\infty^3}\,,\\
  \alpha^{(4)}_{(1,1)}&=
  \frac{9 \pi  \left(5 \gamma ^2-1\right) \theta^{(2)}_{(0,1)} \theta^{(2)}_{(1,0)} v_\infty+12 \pi  \left(7 \gamma ^2-5\right) \gamma ^2 \theta^{(3)}_{(1,1)}-32 \gamma ^2 \theta^{(4)}_{(1,1)} v_\infty}{24 \pi  \gamma ^4 v_\infty^3}\,,\\
  \alpha^{(4)}_{(0,2)}&=
  \frac{9 \pi  \left(5 \gamma ^2-1\right) \delta ^2 (\theta^{(2)}_{(0,1)})^2 v_\infty+24 \pi  \left(7 \gamma ^2-5\right) \gamma ^2 \theta^{(3)}_{(0,2)}-64 \gamma ^2 \theta^{(4)}_{(0,2)} v_\infty}{48 \pi  \gamma ^4 v_\infty^3}\,,\\
  \alpha^{(4)}_{(3,0)}&=
  \frac{v_\infty^2 \theta^{(2)}_{(1,0)}-\theta^{(4)}_{(3,0)}}{4 \gamma  v_\infty^3}\,,\,
  \alpha^{(4)}_{(2,1)}=
  \frac{v_\infty^2 \theta^{(2)}_{(0,1)}-\theta^{(4)}_{(2,1)}}{4 \gamma  v_\infty^3}\,,\,
  \alpha^{(4)}_{(1,2)}=
  -\frac{\theta^{(4)}_{(1,2)}}{4 \gamma  v_\infty^3}\,,\,
  \alpha^{(4)}_{(0,3)}=
  -\frac{\theta^{(4)}_{(0,3)}}{4 \gamma  v_\infty^3}\,.
\end{align}
\end{subequations}
\normalsize
Inserting the scattering-angle coefficients $\theta_{(i,j)}^{(n)}$,
in either the conservative or dissipative, yields the full coefficients.
These are provided in the ancillary file attached to the \texttt{arXiv} submission of this paper 
up to 5PM order, plus a 6PM term at quartic order in spins ($S^4$).

\newpage

\section{Scattering angles}
\label{sec:tables}
\noindent
In this appendix we list the data for the NR simulations, as well, as the scattering-angle predictions for the models considered in this paper. 
We indicate a plunge (or prediction thereof) with a minus, ``$-$''.
We always show the angle in degrees with the fractional uncertainty in 
percentage in brackets.
We note that for the Tables~\ref{Table:AppSpin}, \ref{Table:AppSpinless1} and~\ref{Table:AppSpinless2}, Ref.~\cite{Rettegno:2023ghr} reported an uncertainty in whether the final scattering simulation before plunge, could also have been a plunge.

\begin{table}[b]
  \begin{tabular}{
    |m{3em}
    |m{4em}
    |m{8.3em}|
    |m{8.3em}
    |m{8.3em}
    |m{8.3em}
    |m{8.3em}|
    }
    \hline
    \multicolumn{3}{|c||}{
      \cellcolor{lightgray}
      NR data}&
    \multicolumn{1}{c|}{
      \cellcolor{lightgray}
      SEOB-4PM}&
      \multicolumn{1}{c|}{
        \cellcolor{lightgray}
        $w_{\rm 4PM}$ }&
        \multicolumn{1}{c|}{
          \cellcolor{lightgray}
          SEOB-5PM(I)} &
          \multicolumn{1}{c|}{
            \cellcolor{lightgray}
            $w_{\rm 5PM (I)}$}
    \\
    \hline
    \cellcolor{lightgray} \hfil $\chi_{+}$ &
    \cellcolor{lightgray} \hfil $\Gamma$ &
    \cellcolor{lightgray} \hfil $\theta[{\rm deg}]$ &
    \cellcolor{lightgray} \hfil $\theta[{\rm deg}](\delta \theta[\%])$ &
    \cellcolor{lightgray} \hfil $\theta[{\rm deg}](\delta \theta[\%])$ &
    \cellcolor{lightgray} \hfil $\theta[{\rm deg}](\delta \theta[\%])$ &
    \cellcolor{lightgray} \hfil $\theta[{\rm deg}](\delta \theta[\%])$ 
    \\ 
    \hline \hfil  -0.3 & \hfil 1.02269 & \hfil  $ - $ & \hfil 444.865 & \hfil 521.326 
    &\hfil $-$&\hfil $-$
    \\
    \hline \hfil  -0.25 & \hfil 1.02268 & \hfil  $ 367.545^{+0.}_{-4.84} $ & \hfil 343.778 (-6.5) & \hfil 353.322 (-3.9) & \hfil 404.532 (10.1) & \hfil   $-$ \\
 \hline \hfil -0.23 & \hfil 1.02267 & \hfil $334.345^{+0.084}_{-1.573} $ & \hfil 323.065 (-3.4)  & \hfil 329.103 (-1.6)  & \hfil 357.495 (6.9)  & \hfil 495.03 (48.1)  \\  \hline \hfil -0.22 & \hfil 1.02267 & \hfil $322.693^{+0.099}_{-1.004} $ & \hfil 314.512 (-2.5)  & \hfil 319.468 (-1.)  & \hfil 341.921 (6.)  & \hfil 412.428 (27.8)  \\  \hline \hfil -0.21 & \hfil 1.02267 & \hfil $312.795^{+0.187}_{-0.364} $ & \hfil 306.799 (-1.9)  & \hfil 310.926 (-0.6)  & \hfil 329.068 (5.2)  & \hfil 375.805 (20.1)  \\  \hline \hfil -0.2 & \hfil 1.02266 & \hfil $303.884^{+0.222}_{-0.466} $ & \hfil 299.644 (-1.4)  & \hfil 303.102 (-0.3)  & \hfil 317.95 (4.6)  & \hfil 351.697 (15.7)  \\  \hline \hfil -0.17 & \hfil 1.02266 & \hfil $286.603^{+0.154}_{-0.01} $ & \hfil 281.815 (-1.7)  & \hfil 284.027 (-0.9)  & \hfil 292.662 (2.1)  & \hfil 309.066 (7.8)  \\  \hline \hfil -0.16 & \hfil 1.02266 & \hfil $277.849^{+0.23}_{-0.003} $ & \hfil 276.702 (-0.4)  & \hfil 278.644 (0.3)  & \hfil 285.91 (2.9)  & \hfil 299.349 (7.7)  \\  \hline \hfil -0.15 & \hfil 1.02265 & \hfil $272.603^{+0.26}_{-0.003} $ & \hfil 271.827 (-0.3)  & \hfil 273.537 (0.3)  & \hfil 279.648 (2.6)  & \hfil 290.746 (6.7)  \\  \hline \hfil -0.1 & \hfil 1.02265 & \hfil $251.028^{+0.559}_{-0.003} $ & \hfil 251.762 (0.3)  & \hfil 252.829 (0.7)  & \hfil 255.221 (1.7)  & \hfil 260.003 (3.6)  \\  \hline \hfil -0.05 & \hfil 1.02264 & \hfil $234.568^{+0.845}_{-0.003} $ & \hfil 236.109 (0.7)  & \hfil 236.923 (1.)  & \hfil 237.352 (1.2)  & \hfil 239.441 (2.1)  \\  \hline \hfil 0. & \hfil 1.02264 & \hfil $221.823^{+0.762}_{-0.002} $ & \hfil 223.442 (0.7)  & \hfil 224.179 (1.1)  & \hfil 223.442 (0.7)  & \hfil 224.179 (1.1)  \\  \hline \hfil 0.05 & \hfil 1.02264 & \hfil $211.195^{+0.61}_{-0.002} $ & \hfil 212.83 (0.8)  & \hfil 213.568 (1.1)  & \hfil 212.089 (0.4)  & \hfil 212.078 (0.4)  \\  \hline \hfil 0.1 & \hfil 1.02265 & \hfil $202.608^{+0.388}_{-0.002} $ & \hfil 203.765 (0.6)  & \hfil 204.545 (1.)  & \hfil 202.568 (0.)  & \hfil 202.123 (-0.2)  \\  \hline \hfil 0.15 & \hfil 1.02265 & \hfil $194.542^{+0.183}_{-0.001} $ & \hfil 195.838 (0.7)  & \hfil 196.673 (1.1)  & \hfil 194.359 (-0.1)  & \hfil 193.654 (-0.5)  \\  \hline \hfil 0.2 & \hfil 1.02266 & \hfil $187.838^{+0.02}_{-0.141} $ & \hfil 188.854 (0.5)  & \hfil 189.753 (1.)  & \hfil 187.203 (-0.3)  & \hfil 186.344 (-0.8)  \\  \hline \hfil 0.3 & \hfil 1.02269 & \hfil $176.586^{+0.001}_{-0.653} $ & \hfil 176.997 (0.2)  & \hfil 178.026 (0.8)  & \hfil 175.197 (-0.8)  & \hfil 174.202 (-1.4)  \\  \hline \hfil 0.4 & \hfil 1.02274 & \hfil $167.545^{+0.002}_{-0.941} $ & \hfil 167.228 (-0.2)  & \hfil 168.374 (0.5)  & \hfil 165.423 (-1.3)  & \hfil 164.405 (-1.9)  \\  \hline \hfil 0.6 & \hfil 1.02288 & \hfil $154.139^{+0.005}_{-1.443} $ & \hfil 151.833 (-1.5)  & \hfil 153.156 (-0.6)  & \hfil 150.206 (-2.6)  & \hfil 149.265 (-3.2)  \\  \hline \hfil 0.8 & \hfil 1.02309 & \hfil $145.357^{+0.006}_{-1.528} $ & \hfil 140.053 (-3.6)  & \hfil 141.474 (-2.7)  & \hfil 138.705 (-4.6)  & \hfil 137.879 (-5.1)  \\ 
    \hline
  \end{tabular}
  \caption{
    Spinning NR data of Ref.~\cite{Rettegno:2023ghr} and model predictions.
    The following variables are constant across all data points: $\chi_-=0$ and $\ell=4.5824$.
  }\label{Table:AppSpin}
\end{table}

\end{widetext}
$\hphantom{lalal}$
\newpage
$\hphantom{lalal}$
\begin{table}
  \begin{tabular}{
    |m{3.5em}
    |m{3.1em}
    |m{5.3em}
    ||m{6em}
    |m{6em}|
    }
    \hline
    \multicolumn{3}{|c||}{
      \cellcolor{lightgray}
      NR data}&
      \multicolumn{1}{c|}{
        \cellcolor{lightgray}
        SEOB-4PM}&
        \multicolumn{1}{c|}{
          \cellcolor{lightgray}
          $w_{\rm PM}$}
    \\
    \hline
    \cellcolor{lightgray} \hfil $\Gamma$ &
    \cellcolor{lightgray} \hfil $\ell$ &
    \cellcolor{lightgray} \hfil $\theta_{\rm NR} [{\rm deg}]$ &
    \cellcolor{lightgray} \hfil $\theta[{\rm deg}](\delta\theta(\%))$ &
    \cellcolor{lightgray} \hfil $\theta[{\rm deg}](\delta\theta(\%))$ 
    \\ 
    \hline \hfil 1.00457 & \hfil 4.608 & \hfil $ 201.9^{+4.8}_{-4.8} $ & \hfil 205.227 (1.6)  & \hfil 202.688 (0.4)  \\  \hline \hfil 1.01479 & \hfil 4.6077 & \hfil $ 195.9^{+1.3}_{-1.3} $ & \hfil 197.077 (0.6)  & \hfil 196.008 (0.1)  \\  \hline \hfil 1.01988 & \hfil 4.6076 & \hfil $ 207.03^{+0.99}_{-0.99} $ & \hfil 207.849 (0.4)  & \hfil 207.683 (0.3)  \\  \hline \hfil 1.02496 & \hfil 4.6074 & \hfil $ 225.54^{+0.87}_{-0.87} $ & \hfil 226.256 (0.3)  & \hfil 227.839 (1.)  \\  \hline \hfil 1.03503 & \hfil 4.6061 & \hfil $ 307.13^{+0.88}_{-0.88} $ & \hfil 311.297 (1.4)  & \hfil 334.271 (8.8)  \\ 
    \hline
  \end{tabular}
  \caption{
  Equal-mass, spinless series of NR data from Ref.~\cite{Hopper:2022rwo}.
  The angular momentum is approximately fixed.
  The values of $\Gamma$ and $\ell$ reported here are rounded off.
  For computations, we have used the values reported in Table II of Ref.~\cite{Hopper:2022rwo}.
  The $w_{\rm 4PM}$ angles agree with those reported in Ref.~\cite{Damour:2022ybd}.
}\label{Table:AppVaryE}
\end{table}

\begin{table}
 \begin{tabular}{
    |m{4em}
    |m{6.4em}
    ||m{6.4em}|
    m{6.4em}|
    }
    \hline
    \multicolumn{2}{|c||}{
      \cellcolor{lightgray}
      NR data}&
      \multicolumn{1}{c|}{
        \cellcolor{lightgray}
        SEOB-4PM}&
        \multicolumn{1}{c|}{
          \cellcolor{lightgray}
          $w_{\rm 4PM}$}
    \\
    \hline
    \cellcolor{lightgray}
     \hfil $\ell$ &
    \cellcolor{lightgray}
     \hfil $\theta[{\rm deg}]$ &
     \cellcolor{lightgray} 
     \hfil $\theta[{\rm deg}](\delta \theta[\%])$ &
    \cellcolor{lightgray}
     \hfil $\theta[{\rm deg}](\delta \theta[\%])$
    \\ 
    \hline  \hfil 4.3076 & \hfil $-$ & \hfil 609.322($-$) & \hfil $-$ 
    \\
 \hline \hfil 4.3536 & \hfil $ 376.275^{+0.026}_{-14.69} $ & \hfil 366.188 (-2.7)  & \hfil 379.31 (0.8)  \\  \hline \hfil 4.3764 & \hfil $ 329.057^{+0.003}_{-1.534} $ & \hfil 333.711 (1.4)  & \hfil 341.135 (3.7)  \\  \hline \hfil 4.3808 & \hfil $ 323.422^{+0.}_{-1.914} $ & \hfil 328.694 (1.6)  & \hfil 335.476 (3.7)  \\  \hline \hfil 4.3856 & \hfil $ 318.394^{+0.}_{-1.575} $ & \hfil 323.552 (1.6)  & \hfil 329.728 (3.6)  \\  \hline \hfil 4.39 & \hfil $ 313.764^{+0.}_{-1.331} $ & \hfil 319.111 (1.7)  & \hfil 324.803 (3.5)  \\  \hline \hfil 4.3992 & \hfil $ 305.734^{+0.056}_{-0.694} $ & \hfil 310.548 (1.6)  & \hfil 315.401 (3.2)  \\  \hline \hfil 4.4452 & \hfil $ 274.368^{+0.074}_{-0.016} $ & \hfil 277.789 (1.2)  & \hfil 280.337 (2.2)  \\  \hline \hfil 4.5368 & \hfil $ 235.447^{+0.912}_{-0.003} $ & \hfil 237.465 (0.9)  & \hfil 238.507 (1.3)  \\  \hline \hfil 4.5824 & \hfil $ 221.823^{+0.762}_{-0.002} $ & \hfil 223.442 (0.7)  & \hfil 224.179 (1.1)  \\  \hline \hfil 4.6744 & \hfil $ 200.81^{+0.62}_{-0.004} $ & \hfil 201.581 (0.4)  & \hfil 201.991 (0.6)  \\  \hline \hfil 4.766 & \hfil $ 184.684^{+0.221}_{-0.002} $ & \hfil 185.16 (0.3)  & \hfil 185.411 (0.4)  \\  \hline \hfil 5.0408 & \hfil $ 152.106^{+0.055}_{-0.446} $ & \hfil 152.15 (0.)  & \hfil 152.231 (0.1)  \\  \hline \hfil 5.4992 & \hfil $ 120.804^{+0.013}_{-0.307} $ & \hfil 120.8 (0.)  & \hfil 120.821 (0.)  \\  \hline \hfil 5.9572 & \hfil $ 101.616^{+0.059}_{-0.002} $ & \hfil 101.696 (0.1)  & \hfil 101.704 (0.1)  \\  \hline \hfil 6.4156 & \hfil $ 88.26^{+0.337}_{-0.002} $ & \hfil 88.42 (0.2)  & \hfil 88.423 (0.2)  \\  \hline \hfil 6.874 & \hfil $ 78.296^{+0.52}_{-0.002} $ & \hfil 78.514 (0.3)  & \hfil 78.515 (0.3)  \\  \hline \hfil 7.332 & \hfil $ 70.404^{+0.927}_{-0.003} $ & \hfil 70.776 (0.5)  & \hfil 70.777 (0.5)  \\ 
    \hline
  \end{tabular}
  \caption{
    First equal-mass, spinless series of NR data from Ref.~\cite{Rettegno:2023ghr} and model predictions.
    The $w_{\rm 4PM}$ predictions computed here are identical to the ones given in Ref.~\cite{Rettegno:2023ghr}.
    The energy is fixed with $\Gamma=1.02264$.
  }\label{Table:AppSpinless1}
\end{table}

\begin{table}
  \begin{tabular}{
    |m{4em}
    |m{6.4em}
    ||m{6.4em}|
    m{6.4em}|
    }
    \hline
    \multicolumn{2}{|c||}{
      \cellcolor{lightgray}
      NR data}&
      \multicolumn{1}{c|}{
        \cellcolor{lightgray}
        SEOB-4PM}&
        \multicolumn{1}{c|}{
          \cellcolor{lightgray}
          $w_{\rm 4PM}$}
    \\
    \hline
    \cellcolor{lightgray} \hfil $\ell$ &
    \cellcolor{lightgray} \hfil $\theta[{\rm deg}]$ &
    \cellcolor{lightgray} \hfil $\theta[{\rm deg}](\delta\theta(\%))$ &
    \cellcolor{lightgray} \hfil $\theta[{\rm deg}](\delta\theta(\%))$ 
    \\ 
    \hline \hfil 4.602 & \hfil $-$ & \hfil $-$ & \hfil $-$  \\
    \hline \hfil 4.638 & \hfil $ 392.815^{+0.006}_{-7.477} $ & \hfil 413.403 (5.2)  & \hfil $-$  \\  \hline \hfil 4.662 & \hfil $ 338.973^{+0.156}_{-0.756} $ & \hfil 359.875 (6.2)  & \hfil 430.914 (27.1)  \\  \hline \hfil 4.68 & \hfil $ 317.637^{+0.142}_{-0.444} $ & \hfil 334.244 (5.2)  & \hfil 374.716 (18.)  \\  \hline \hfil 4.722 & \hfil $ 283.359^{+0.343}_{-0.007} $ & \hfil 294.155 (3.8)  & \hfil 312.437 (10.3)  \\  \hline \hfil 4.758 & \hfil $ 262.825^{+0.749}_{-0.008} $ & \hfil 270.784 (3.)  & \hfil 282.302 (7.4)  \\  \hline \hfil 4.8 & \hfil $ 244.21^{+1.22}_{-0.005} $ & \hfil 250.183 (2.4)  & \hfil 257.74 (5.5)  \\  \hline \hfil 5.04 & \hfil $ 184.138^{+0.439}_{-0.004} $ & \hfil 186.039 (1.)  & \hfil 187.734 (2.)  \\  \hline \hfil 5.28 & \hfil $ 153.119^{+0.226}_{-0.227} $ & \hfil 153.934 (0.5)  & \hfil 154.603 (1.)  \\  \hline \hfil 5.4 & \hfil $ 141.986^{+0.244}_{-0.213} $ & \hfil 142.634 (0.5)  & \hfil 143.094 (0.8)  \\  \hline \hfil 5.64 & \hfil $ 124.805^{+0.154}_{-0.238} $ & \hfil 125.244 (0.4)  & \hfil 125.487 (0.5)  \\  \hline \hfil 5.7 & \hfil $ 121.233^{+0.18}_{-0.153} $ & \hfil 121.669 (0.4)  & \hfil 121.88 (0.5)  \\  \hline \hfil 5.76 & \hfil $ 117.897^{+0.157}_{-0.091} $ & \hfil 118.333 (0.4)  & \hfil 118.517 (0.5)  \\  \hline \hfil 6. & \hfil $ 106.459^{+0.207}_{-0.004} $ & \hfil 106.904 (0.4)  & \hfil 107.016 (0.5)  \\  \hline \hfil 7.2 & \hfil $ 73.095^{+1.358}_{-0.006} $ & \hfil 73.808 (1.)  & \hfil 73.826 (1.)  \\  \hline \hfil 8.4 & \hfil $ 56.489^{+1.242}_{-0.006} $ & \hfil 57.155 (1.2)  & \hfil 57.16 (1.2)  \\  \hline \hfil 9.6 & \hfil $ 45.982^{+1.53}_{-0.008} $ & \hfil 46.866 (1.9)  & \hfil 46.868 (1.9)  \\ 
    \hline
  \end{tabular}
  \caption{
    Second equal-mass, spinless series of NR data from Ref.~\cite{Rettegno:2023ghr} and model predictions.
    The $w_{\rm 4PM}$ predictions computed here are identical to the ones given in Ref.~\cite{Rettegno:2023ghr}.
    The energy is fixed with $\Gamma=1.04033$.
  }\label{Table:AppSpinless2}
\end{table}

\begin{table}
  \begin{tabular}{
    |m{4em}
    |m{6.4em}
    ||m{6.4em}|
    m{6.4em}|
    }
    \hline
    \multicolumn{2}{|c||}{
      \cellcolor{lightgray}
      NR data}&
      \multicolumn{1}{c|}{
        \cellcolor{lightgray}
        SEOB-4PM}&
        \multicolumn{1}{c|}{
          \cellcolor{lightgray}
          $w_{\rm 4PM}$}
    \\
    \hline
    \cellcolor{lightgray} \hfil $\ell$ &
    \cellcolor{lightgray} \hfil $\theta[{\rm deg}]$ &
    \cellcolor{lightgray} \hfil $\theta[{\rm deg}](\delta\theta(\%))$ &
    \cellcolor{lightgray} \hfil $\theta[{\rm deg}](\delta\theta(\%))$ 
    \\ 
    \hline  \hfil 4.2 & \hfil $-$ & \hfil $-$ & \hfil $-$ \\
    \hline \hfil 4.9 & \hfil $ 354.118^{+0.307}_{-0.633} $ & \hfil $-$ & \hfil $-$  \\  \hline \hfil 5.04 & \hfil $ 248.95^{+1.203}_{-0.005} $ & \hfil 273.639 (9.9)  & \hfil 280.445 (12.7)  \\  \hline \hfil 5.18 & \hfil $ 206.064^{+1.479}_{-0.006} $ & \hfil 216.687 (5.2)  & \hfil 218.732 (6.1)  \\  \hline \hfil 5.32 & \hfil $ 179.815^{+0.484}_{-0.006} $ & \hfil 185.602 (3.2)  & \hfil 186.552 (3.7)  \\  \hline \hfil 5.6 & \hfil $ 146.516^{+0.354}_{-0.096} $ & \hfil 149.089 (1.8)  & \hfil 149.418 (2.)  \\  \hline \hfil 6.3 & \hfil $ 104.166^{+0.361}_{-0.006} $ & \hfil 105.225 (1.)  & \hfil 105.287 (1.1)  \\  \hline \hfil 7. & \hfil $ 82.275^{+0.924}_{-0.007} $ & \hfil 83.171 (1.1)  & \hfil 83.191 (1.1)  \\  \hline \hfil 7.7 & \hfil $ 68.351^{+1.485}_{-0.007} $ & \hfil 69.33 (1.4)  & \hfil 69.339 (1.4)  \\ 
    \hline
  \end{tabular}
  \caption{
    Third equal-mass, spinless series of NR data from Ref.~\cite{Rettegno:2023ghr} and model predictions.
    The $w_{\rm 4PM}$ predictions computed here are identical to the ones given in Ref.~\cite{Rettegno:2023ghr}.
    The energy is fixed with $\Gamma=1.05548$.
  }\label{Table:AppSpinless3}
\end{table}

\clearpage

\bibliographystyle{JHEP}
\bibliography{../wqft_spin}

\end{document}